\documentclass[acmsmall]{acmart}
\usepackage{lscape}
\usepackage{longtable}
\usepackage{multirow}
\usepackage[utf8]{inputenc}
\usepackage[T2A,T1]{fontenc}
\usepackage[english,russian]{babel}

\AtBeginDocument{%
  }

\begin{document}

\title{Digital gazetteers: review and prospects for place name knowledge bases}

\author{Kalana Wijegunarathna}
\email{k.wijegunarathna@massey.ac.nz}
\orcid{0000-0001-7458-4801}
\author{Kristin Stock}
\email{k.stock@massey.ac.nz}
\orcid{0000-0002-5828-6430}
\affiliation{%
  \institution{School of Mathematical and Computational Sciences, Massey University}
  \city{Auckland}
  \country{New Zealand}
}

\author{Christopher B. Jones}
\email{jonescb2@cardiff.ac.uk}
\orcid{0000-0001-6847-7575}
\affiliation{%
  \institution{School of Computer Science and Informatics, Cardiff University}
  \city{Cardiff}
  \country{UK}
}









\begin{abstract}
 Gazetteers typically store data on place names, place types and the associated coordinates. They play an essential role in disambiguating place names in online geographical information retrieval systems for navigation and mapping,  detecting and disambiguating place names in text, and providing coordinates. Currently there are many gazetteers in use derived from many sources, with no commonly accepted standard for encoding the data. Most gazetteers are also very limited in the extent to which they represent the multiple facets of the named places yet they have potential to assist user search for locations with specific physical, commercial, social or cultural characteristics.  With a view to understanding digital gazetteer technologies and advancing their future effectiveness for information retrieval, we provide a review of data sources, components, software and data management technologies, data quality and volunteered data, and methods for matching sources that refer to the same real-world places. We highlight the need for future work on richer representation of named places, the temporal evolution of place identity and location, and the development of more effective methods for data integration. 
  
\end{abstract}




\maketitle

\section{Introduction}
Gazetteers have a long history of use to represent information about places. Digital gazetteers usually take the form of databases or knowledgebases that store information about place names (also referred to as toponyms) and support access to many geospatial applications. They may also be referred to as geographical gazetteers to distinguish them from the less common use of the term gazetteer in natural language processing, to refer to a dictionary or a list of names that includes places, but also organizations, people, and domain-specific terms \cite{shaalan2008arabic,nguyen2014next}. The earliest geographical gazetteers{\footnote{The word gazetteer dates back at least to the seventeenth century, to describe either a journalist or a newspaper, while its use to refer to a geographic index of places became more common in the nineteenth century (Oxford English Dictionary https://www.oed.com/).}} listed place names and described places with regard to their physical and social characteristics. Ptolemy's \textit{Geographia} appears to be one of the earliest recorded gazetteers (although it is a revision of a now lost atlas believed to have predated it), and was primarily used for navigation \cite{berggren2000ptolemy}. Digital gazetteers tend to be considerably less comprehensive than historical gazetteers, often lacking extensive place descriptions (see the Gazetteer for Scotland\footnote{https://www.scottish-places.info/} for a rare exception). However, they almost always include geographical coordinates (e.g., latitude and longitude) along with the type of place (e.g. river, town), referred to in this paper as geographic feature type. Additional elements can include alternative place names, demographic information, and details of the place name etymology. With this shift, gazetteers are now often used to support place name-based search, such as in web mapping and navigation, or to support geospatial referencing (geoparsing and georeferencing) of text documents \cite{hill2000core}.

Gazetteers have become a cross-cutting research topic encompassing Geographic Information Science (GIS), Information Retrieval (IR), computer science, and history. They play a key role in systems and research that exploit or depend upon the use of place names to access or manage information. For example, in scenarios involving web or social media searches, web mapping services, or digital library queries aimed at locating geographic information, a gazetteer could be employed to recognise and disambiguate (find the correct instance of) a place name in the query, and use the resulting coordinates to  pan and zoom to the relevant location on a map \cite{buckland2007geographic, cai2002geovsm, egenhofer2002toward, purves2018geographic, riekert2002automated, souza2005role, sultanik2012rapid, tochtermann1997using, weaver2003digital}. Routing services for emergency response, public transport or navigation could also employ gazetteers to recognise user-specified place names. The utilization of social media content for disaster response may require a gazetteer to locate named places in the disaster zone. The processes of toponym recognition and toponym resolution (disambiguation and geocoding) are relevant both to understanding a spatial query and to analysing text to find geographical references, referred to as georeferencing. The presence of a candidate name in a gazetteer can be part of the evidence for identification as a toponym, while the disambiguation and geocoding process can use gazetteers to generate their coordinates, for example \cite{lieberman2010geotagging}. Current named entity recognition (NER)  methods, of which toponym recognition is a sub-task, may dispense with a gazetteer, relying instead on learning from expansive corpora. Although language modelling methods have been developed to generate coordinates for toponyms or text documents, a gazetteer remains advantageous for generating precisely disambiguated coordinates \cite{melo2017automated, delozier2015gazetteer}.

In addition to the above applications, gazetteers have been developed for applications in the humanities, for example to record names and locations of historical and archaeological artefacts and places of interest. An example is The World Historical Gazetteer (WHG) which can be searched temporally and is contributed to by archaeologists and historical researchers \cite{grossner2021linked}. Temporal search of gazetteers can support study of the evolution of names \cite{burenhult2008language, derungs2013meanings, yoshikatsu2017geographic}. Other history-related applications of gazetteers include \cite{blank2015geocoding, horne2020beyond, lin2020displaying, zhang2020contemporary}. Gazetteers also act as a source for ethnophysiographic studies of how languages and cultures refer to natural landscape terms \cite{mark2003ethnophysiography}.

Transformations in computer science, information retrieval and technology have led to digital gazetteers evolving in some cases from being simple name lookups to becoming aspects of complex geospatial knowledgebases such as Google Maps\footnote{https://www.google.com/maps} and OpenStreetMap (OSM) \cite{ballatore2013survey} that combine typical gazetteer functionality with storage of detailed digital map data. Many of the developments have led however to experimental gazetteers produced by researchers whose work has not been properly integrated or maintained. Notably, most widely used gazetteers have no common content standard or feature type schema - and the call \cite{kessler2009agenda} for an agreed feature type ontology has not been met. The result is a proliferation of gazetteers which, while potentially regionally comprehensive, have limited or no alignment with respect to whether their respective records refer to the same place, or with regard to access methods.  Whether it is possible or even desirable to follow a single standard for gazetteers is also a question that has been raised, due to the wide variety of both uses and users for gazetteers \cite{hill2006gazetteers, kessler2009agenda}. Understanding the diversity of approaches to gazetteer design and potential for greater integration of gazetteer content is one of the main motivations behind our review of gazetteers, gazetteer technologies and standards.

While digital gazetteers have a wide range of applications, there is often little information about the places referred to by toponyms. This is in contrast to traditional pre-digital gazetteers that, as indicated above, often recorded a much broader range of information about a place including for example its historical origins, events that have taken place there, its architecture and types of services that it affords. The latter could include religious institutions, commercial companies and industrial factories. Numerous academic publications have considered the multiple dimensions or facets of the concept of place, including how they can be perceived by different people and hence, by implication, why there could be multiple reasons for particular places to be of interest \cite{tuan1977space, Purves-Place-2019, Hamzei2020PlaceFacets}. There is a motivation therefore to maintain such facets in a digital gazetteer.

When people query web mapping systems they frequently combine a place name with a service or activity of interest, perhaps relating to food and drink, leisure activities, culture, sport, or some commercial enterprise. Given that the main general purpose digital gazetteers provide either no or limited support for the service and affordance aspects of such queries, the search engines can be expected to depend upon curated data sources in the form of point of interest (POI) datasets, or on web pages that associate a place name with a service or activity. A major source of more substantial data about places is provided by the place-specific pages in Wikipedia and their semantic web presentations on DBpedia, but there is great variation in the content and geographic coverage of these pages. In conventional digital gazetteers the place type data item enables some description of the named place, but often this is recorded as only a single feature type. It may be argued that gazetteers could provide improved support for geospatial queries, and serve as more general purpose knowledge bases of named places, if they recorded much richer information about the place to which the name refers. The place could then be regarded as a multi-faceted information object that is the basis of a gazetteer record \cite{Purves-Place-2019}.

In this paper, we review publications addressing the topic of gazetteers published since the last decade of the 0\textsuperscript{th} century. We draw comparisons between papers through various aspects: reviewing the current standing of the standards for gazetteers; assessing the numerous methods for identifying, deduplicating and disambiguating place names; integrating place name sources; identifying gaps and potential for improvement. We also review the data items commonly found in gazetteers, and we highlight the distinction, indicated above, between gazetteers as quite sparse representations of place name knowledge and the potential for gazetteers to reinstate more widely their standing as rich representations of knowledge about the named places. Our intention is to provide insights that will help future gazetteer developers make decisions on these areas. We also propose areas for further research and areas that may benefit from the integration of the latest developments in computer science, particularly those of deep learning and AI.

We will address these research questions in this paper: 
\begin{enumerate}
    \item What are the core components of a gazetteer and how have they evolved over the years?
    \item What place name data sources have been used to populate gazetteers?
    \item What methods have been most effective in the compilation and integration of gazetteers?
    \item What implications has the growth of the web and Volunteered Geographic Information (VGI) had on the building of gazetteers?
    \item What technologies have been used for the implementation of gazetteers?
    \item What are the limitations of digital gazetteers and how can they be addressed?
\end{enumerate}

The remainder of the paper is structured as follows: In the next section, we summarise previous relevant reviews in the area. In Section 3, we present the methodology for the literature review. In Section 4, we describe the evolution and classification of gazetteers, followed by a discussion of the contents of a gazetteer in Section 5. In Section 6, we discuss the sources used to build gazetteers and the process of integration of these sources. Section 7 discusses VGI and gazetteers on the web, and Section 8 reviews gazetteer technologies. In Section 9, we discuss the improvements that can be made to gazetteers to suit modern applications and avenues for future research, while Section 10 sumarises conclusions.

\section{Related Studies}

Upon conducting an extensive search, we were unable to identify publications primarily focused on reviewing the existing literature on digital gazetteers. Nonetheless, several papers that address the broad topic of the construction and evolution of gazetteers were found, some of which consist of a substantial review component. These findings are summarized here.

\cite{hill2000core} discusses gazetteers in light of the Alexandria Digital Library (ADL) gazetteer. A gazetteer is defined as a geospatial dictionary of geographic names with the core components of a name, a location, and a type. This definition is one of the most frequently cited of a modern gazetteer. The paper discusses the state of gazetteers at that point in time, elaborating on the three main components, with particular regard to the ADL, and discussing underlying principles relevant to place names. The paper provides an informative introduction to gazetteers in the late 20\textsuperscript{th} century.

In proposing future directions and challenges for gazetteers, \cite{kessler2009agenda} identify shortcomings of existing gazetteers and upcoming trends, particularly the incorporation of VGI, discussing challenges and solutions. A more recent article on gazetteers \cite{goodchild2016gazpresentspatial} discusses the prominence of VGI as a key component of \textit{neogeography} \cite{turner2006introduction}, the phenomenon of people, particularly outside the discipline of geography, creating and interacting with digital geographic information and maps. Even though their interest is not in reviewing gazetteers, they do provide an introduction to gazetteers. 

\cite{polczynski2022lessons} aim to guide new gazetteer builders in making design decisions based on the authors' experience. Though they discuss the merits of particular approaches to building and publishing gazetteers (such as Linked Open Data), they do not present a current standing of gazetteers or study their components in depth. 
 
None of these papers has comprehensively discussed the evolution of gazetteers, their components, and the application of latest technologies in constructing a gazetteer, as we aim to here.

\section{Methodology}

In order to answer our research questions, we designed two search queries to run on three main source databases: Scopus, Web of Science and EBSCO Discover. The search queries are as follows:

\begin{enumerate}
    \item (TITLE(gazetteer*) AND TITLE-ABSTRACT-KEYWORDS(digital OR geograph*)) OR (TITLE(place names) AND TITLE-ABSTRACT-KEYWORDS(gazetteer*))
    \item TITLE (place name database OR place name ontology OR place name knowledge base OR place name knowledgebase OR toponym database OR toponym ontology OR toponym knowledgebase OR toponym knowledge base)
\end{enumerate}

The first query was crafted to encompass as many papers as possible while limiting the focus to geographical gazetteers. However, this approach may have inadvertently excluded papers where gazetteers were employed but referred to by alternative terminology, or were included within wider knowledge bases, ontologies, or databases, and hence the second query was also included. Both queries were limited to journal papers and peer reviewed conference papers published from the year 2000 to 2022. We then enhanced the searches with papers identified through examination of the reference lists of relevant papers. This helped us to capture papers that were not directly retrieved from our search queries including some before the year 2000.  Among papers resulting from our initial search queries, we came across some that were not directly relevant to the topic or answered any of the RQs. Therefore, we excluded papers that met the following exclusion criteria:

\begin{enumerate}
    \item Papers that did not use gazetteers.
    \item Papers discussing older gazetteers which were not exclusively geographic but more administrative catalogs.
    \item Papers that used the term "gazetteer" to refer to dictionaries or name records (which may or may not include place names).
    \item Papers that publish simple lists of place names
\end{enumerate}


A breakdown of the number of papers extracted is shown in Appendix A Figure 1.  


A few other references relevant to our research questions are included, which were not in the search results or cited by them, but which the authors were aware of or were found through searches using Google Scholar. Google Scholar was not included as a source in the first search due to the broad range of materials that it includes that may not be peer-reviewed.

\section{The Evolution and Classification of Gazetteers}

\begin{figure}[H]
\centering 
\includegraphics[width=0.4\textwidth]{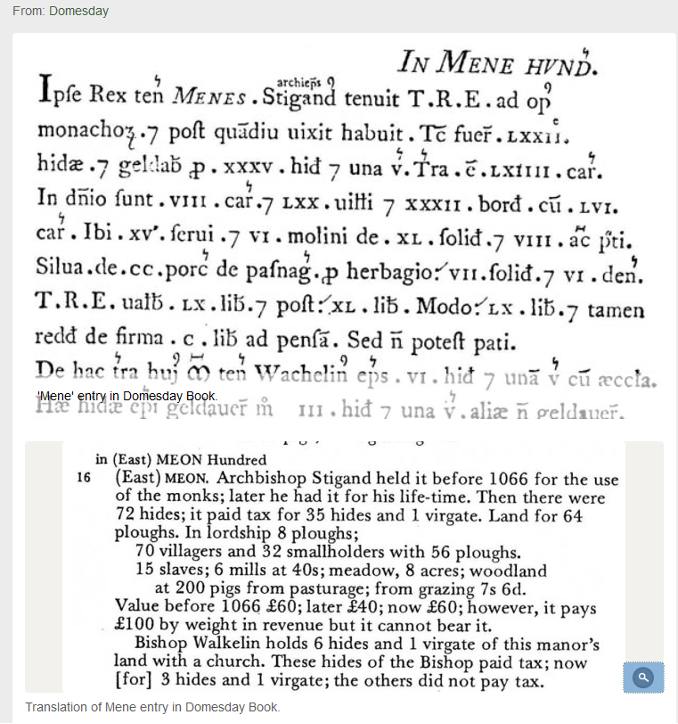}
    \caption[The system.]{Example of content from Domesday Book (Source: https://www.eastmeonhistory.org.uk/content \ \ /catalogue\_item/domesday/mene-domesday-book. Translated by Gordon Timmins). }
\label{Fig2}
\end{figure}

Many historical gazetteers were created by Kings, Emperors or rulers of kingdoms and contained more information than is commonly found in contemporary gazetteers. Though often lacking coordinate pairs or other forms of geographic footprints, these gazetteers commonly include information such as population, descriptions of the boundaries of the place, landmarks that made the identification of the place easier, historically well-known people or incidents, and administrative information such as records of tax collection and commodities \cite{a4575054-6aeb-30da-b0a0-f1d85691d66e}. For example, the Domesday Book recorded an extensive survey of Britain in 1086, listing land area; number of people (including slaves); livestock; income and taxation (see Fig. \ref{Fig2}). From the late eighteenth century, similar forms of gazetteers became increasingly common, sometimes also including local information about facilities such as train stations and schools (see for example Wilson's Gazetteer of Scotland \cite{gazofscot1882Wilson}).

In the mid twentieth century, gazetteers adopted a more quantitative approach to spatial information, focusing on collection and storage of coordinates or other spatial footprints, feature types and administrative divisions that allowed the location of a place to be specified accurately \cite{d1f217a9-23c2-3411-8f2a-e3588c408c9c}. Historically, gazetteers were primarily curated by national mapping authorities and detailed the place names within a specific country. The first initiative to compile a global gazetteer by aggregating individual national gazetteers was undertaken by the Economic and Social Council of the United Nations (ECOSOC) in the 1950s, although the concept was initially proposed at the Fifth International Geographical Congress in 1891. In the 1960s and 70s, the United Nations Group of Experts on Geographical Names (UNGEGN) passed several resolutions with various recommendations about definitions, standards and resolution of national differences. However, the envisaged world gazetteer did not come to fruition due to differences among countries \cite{rsingh2016InternationalStandards}. Nevertheless, initiatives taken by individual countries became imperative to the development of various gazetteer-like services in the last few decades of the twentieth century \cite{d1f217a9-23c2-3411-8f2a-e3588c408c9c}. 

\begin{figure}[H]
\centering 
\includegraphics[width=0.7\textwidth]{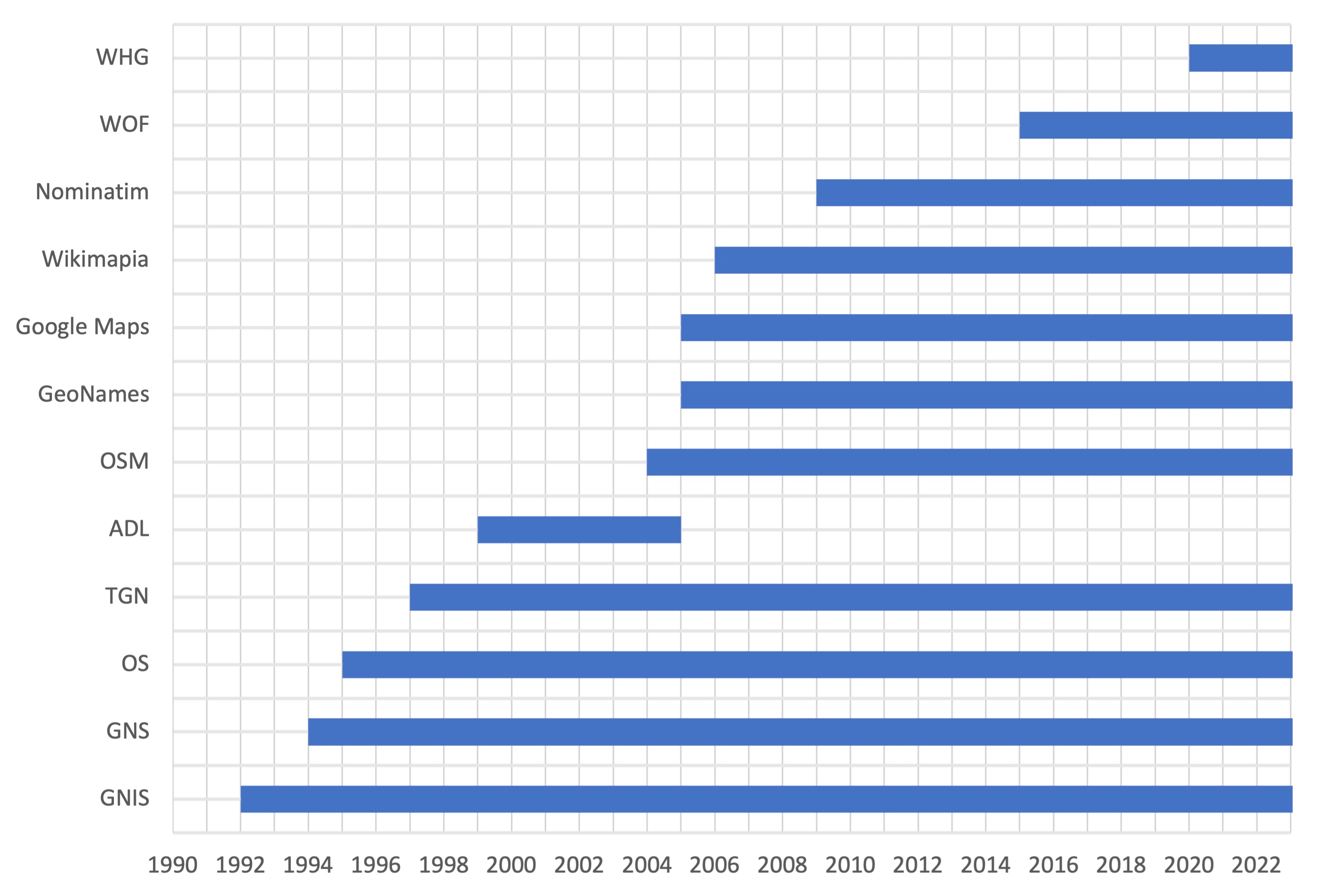}
    \caption[The system.]{Timeline of Commonly Referenced Gazetteers. (see text for acronyms)}
\label{Fig3}
\end{figure}

Fig. \ref{Fig3} shows a timeline for some well-known, commonly used gazetteers. Not all were returned by our search queries (as they are internet applications), but they are examples of popular gazetteers (or services that act as gazetteers even though named otherwise). The list is not exhaustive. All of them exist or have existed as commercial or publicly available place name information sources. 

The Geographic Names Information System (GNIS)\footnote{https://www.usgs.gov/tools/geographic-names-information-system-gnis}  is a place name database managed by the U.S. Board on Geographic Names (BGN)\footnote{https://www.usgs.gov/us-board-on-geographic-names}, the authoritative body responsible for place names in the U.S. The GeoNET Names Server (GNS)\footnote{https://geonames.nga.mil/geonames/GeographicNamesSearch/}  was created by the U.S. National Geospatial-Intelligence Agency (formerly the National Imagery and Mapping Agency) and was originally responsible for collecting the names of places outside the U.S. These gazetteers are listed as the sources for the Alexandria Digital Library Gazetteer (ADL). Even though the ADL gazetteer was only in operation for a short period of time, the developers attempted to build a common and comprehensive gazetteer standard known as the Gazetteer Content Standard (GCS), along with the ADL Feature Type Thesaurus (FTT)  type scheme for places (see Section 5.2). The GCS has not been adopted as a de facto standard, but it incorporates a wide range of data items including alternative names, temporality, multiple geometries, pronunciation and etymology of names, certainty of attributes and provenance information. Due to the scarcity of data, a significant number of records in the ADL gazetteer lacked many of the proposed GCS attributes. We discuss the GCS, the FTT and the ADL gazetteer in light of specific topics throughout the paper.

Britain's Ordnance Survey (OS)\footnote{https://www.ordnancesurvey.co.uk/}  completed the digitization of all its maps in 1995 along with an associated gazetteer. It was accompanied by the development of many digital gazetteers listing property addresses managed by local government agencies. These property and street gazetteers were subsequently managed jointly by a centralized agency GeoPlace\footnote{https://www.geoplace.co.uk}. The British Ordnance Survey is just one of many national mapping agencies that have produced gazetteers or place name databases in association with their map products. The Getty Thesaurus of Geographic Names (TGN) \footnote{https://www.getty.edu/research/tools/vocabularies/tgn/index.html} was compiled and maintained by the Getty Research Institute. It does not call itself a gazetteer, but it is a place name knowledge base with many geographic footprints \cite{hill2000core, hu2019natural}.

The early twenty-first century saw a prevalence of open gazetteers that use publicly volunteered (crowdsourced) information, including OpenStreetMap (OSM)\footnote{https://www.openstreetmap.org/}, GeoNames\footnote{https://www.geonames.org/} and Wikimapia\footnote{https://wikimapia.org/}. OpenStreetMap is associated with the Nominatim\footnote{https://nominatim.org/} gazetteer tool to access the place names in the OSM database. Google Maps is a large commercial undertaking launched in 2005 and Google Places is a service that provides place name information. These products that had their inception in the early 2000s now have significant coverage of the earth with millions of data points. Who's on First (WOF)\footnote{https://whosonfirst.org/} is a recent project to build a global gazetteer. The World Historical Gazetteer (WHG)\footnote{https://whgazetteer.org/} is a gazetteer listing historical places to aid historians and other humanities researchers to map historical places, artefacts, and events.

\begin{figure}
\centering 
\includegraphics[width=0.7\textwidth]{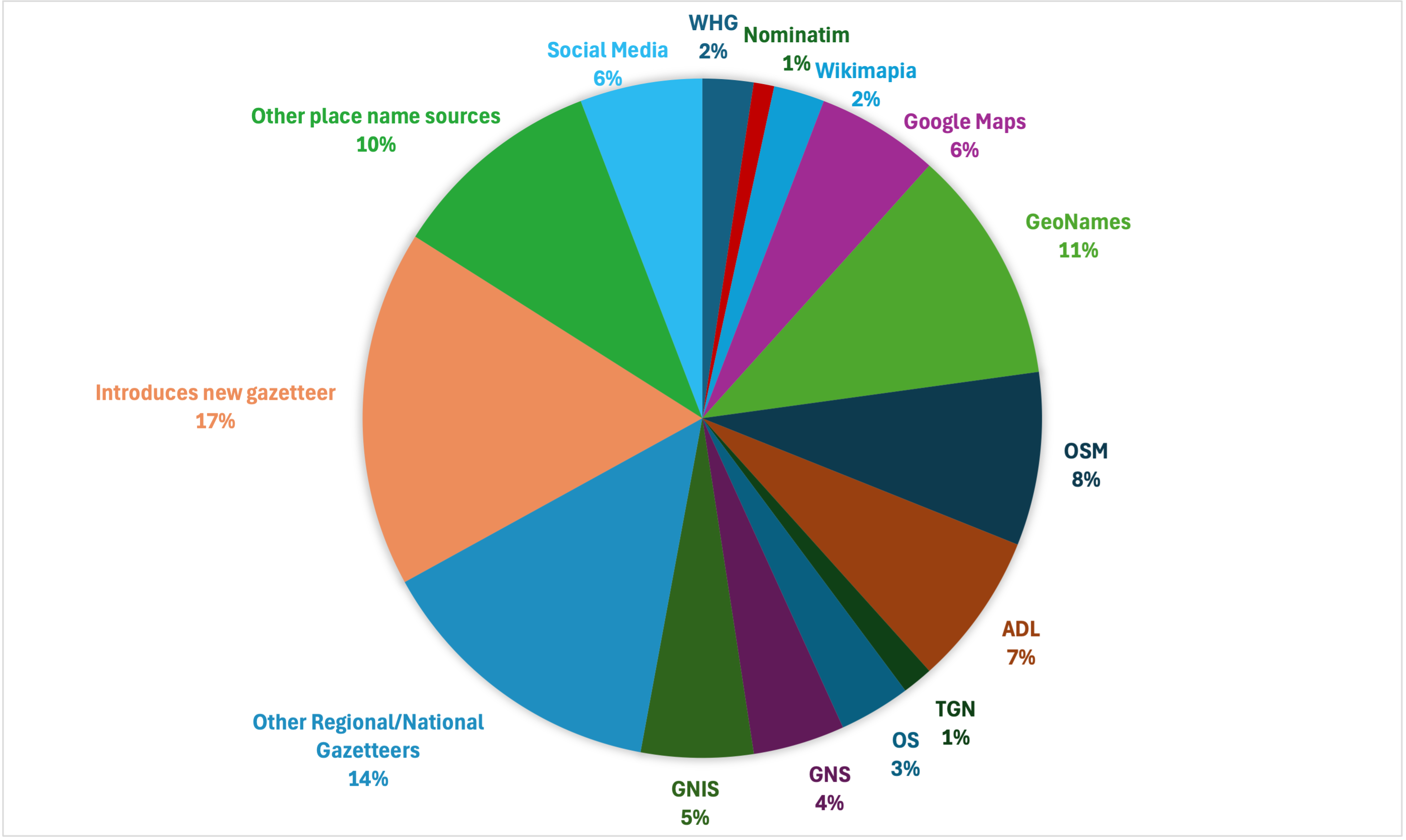}
    \caption[The system.]{Usage of the prominent gazetteers in retrieved articles. Figure also presents articles that introduce new gazetteers and source non-gazetteer sites as place name information sources.}
\label{Fig35}
\end{figure} 

We present the frequency of usage of these prominent gazetteers in the articles retrieved in Fig. \ref{Fig35}. In addition to the application of existing gazetteers, we also report here the articles that introduce new gazetteers, and articles that harvest various other sites, notably social media, as place name sources (e.g. Flickr, Panoramio, Twitter). Other non-gazetteer resources that have been treated or harvested as place knowledgebases (e.g. Wikipedia, bio-diversity collection records) are recorded under the "Other place name sources" category. While some prominent and more frequently used national gazetteers, such as the British OS and GNIS, have been included separately, many other national or regional gazetteers were identified (e.g. SwissTopo, China Historical GIS (CHGIS), Norwegian Central Place Name Register (SSR), Chinese National Geomatics Center's gazetteer). This category  of "Other Regional/National Gazetteers" also includes ancient gazetteers which are almost always regional. "Google Maps" includes both Google Maps and the Google Places API. In the context of this figure, the use of a gazetteer could be to create a new gazetteer, the use of places from a gazetteer for the creation of datasets to test (and or train) new models or methods, or an analysis of a gazetteer (e.g. articles discussing the coverage or data quality of specific gazetteers).

Free and crowdsourced gazetteers like GeoNames and OSM are the most commonly used established gazetteers. This observation is quite intuitive given their easy accessibility, but also indicates the level of confidence placed in crowdsourced data and the measures taken by these gazetteers to manage the quality of VGI. Another observation of interest is the absence of any papers (among those retrieved) that conducted research using Google Maps/Google Places API since 2018, when Google introduced a significant overhaul to its Maps Platform pricing model, making it no longer freely available for developers. Despite being discontinued, the ADL gazetteer remains a prominent knowledge source not only as a gazetteer itself (when it was available) but for the ADL GCS. This reflects its original intentions for building a common gazetteer standard and data model. The large number of gazetteers that are introduced in the papers, however, hints at the acute challenge of defining and gaining adoption for such a standard. 

Over the years the feature types of places in gazetteers have been encoded with various knowledge representation schemes in the semantic spectrum. The semantic spectrum \cite{obrst2010ontologySpectrum} ranks different semantic technologies according to their semantic strength and interoperability. Even though according to Obrst \cite{obrst2010ontologySpectrum}, the semantic spectrum ranks ontologies, it often also includes simpler and less semantically powerful Knowledge Organization Systems (KOS) like lists, dictionaries, and controlled vocabularies \cite{obrst2010ontologySpectrum}. It is beyond the scope of this paper to discuss the semantic spectrum in detail, but it is vital to understand some of the knowledge organisation systems used in gazetteers. We will discuss feature typology in detail in Section 5 but introduce some KOS here since they play a vital part in the classification of gazetteers. Some gazetteers use these systems for defining place types while there are others that use them as their data structure to maintain all contained place names.

The simplest models are controlled vocabularies - simply a finite collection of items. A glossary is slightly more advanced using free text to describe the meanings of terms. Increasing semantic interoperability leads to thesauri and taxonomies. A thesaurus provides synonyms, broader/narrower terms and association relations for its terms. Taxonomies provide stronger hierarchical relations in addition to the features provided by a thesaurus. More semantically enriched are conceptual models and logical theories. A conceptual model is considered a weak ontology and a logical theory is a strong one \cite{obrst2010ontologySpectrum}, where an ontology is defined as an explicit specification of a conceptualization \cite{gruber1995toward}. A conceptualization is made up of a set of objects, concepts, and relations between them about which knowledge is being expressed. A conceptual model has generalised relations (class level), properties, instances, and attributes, while a logical theory is enhanced further with axioms and rules. Logical theory enables machine semantic interpretation as it is represented in a logical knowledge representation. Both these types of ontologies can have concept attributes, and different types of relations.

Gazetteers use different semantic models from various levels in the spectrum. Early digital gazetteers used controlled vocabularies and thesauri (GNIS, GNS, ADL, GeoNames). The beginning of the 21st century saw a trend toward use of web-based ontologies, discussed in the section on Gazetteer Technologies. In parallel with the  use of ontologies was the use of  semantically enriched typology exemplified by folksonomies. Folksonomies can be regarded as a type of taxonomy where the public tags online items. The folksonomy used by OSM enforces a loose hierarchy but users or the public who add a place to OSM are allowed to tag the places with their own place types - leading to a large number of tags whose similarities or differences, even within the same hierarchy are sometimes hardly discernible. We discuss the use of feature type models in detail in Sections \ref {sec:FeatureType} and \ref{sec:GazTechs}.

\section{Components of a Gazetteer}
\label{sec:Components}
The three main components of a gazetteer identified earlier: place name, feature type and geographic footprint, have remained the backbone of modern gazetteers. Depending on the gazetteer's use cases, various other changes and additions have been proposed to make them more useful along with various extensions to the three main components.

\subsection{Place Names}

Place names are an essential component of a gazetteer as they provide the most common means by which people refer to locations. The use of place names however introduces challenges of ambiguity, duplication, multilingualism and the need to resolve and integrate local and vernacular place names.  \cite{laurini2015geographic} discusses these challenges referring to each of the issues we present with examples in Table 1.

\begin{table}[]
\small
\caption{Challenges in dealing with place names.}

\begin{tabular}{lll}
\hline
\textbf{Challenge}                                                           & \textbf{Description}                                                                                                                                                                                                                                                                             & \textbf{Example}                                                                                                                                                                                                                                                                \\ \hline
\textbf{\begin{tabular}[c]{@{}l@{}}Feature \\ ambiguity\end{tabular}}        & \begin{tabular}[c]{@{}l@{}}The same name could refer to two different \\ features because they are of two different \\ feature types.\end{tabular}                                                                                                                                               & \begin{tabular}[c]{@{}l@{}}New Zealand, the country, is \\ different from New Zealand the \\ group of islands referring to a \\ physical entity.\end{tabular}                                                                                                                   \\ \hline 
\textbf{Multilingualism}                                                     & \begin{tabular}[c]{@{}l@{}}The same place being called different names \\ in different languages.\end{tabular}                                                                                                                                                                                   & \begin{tabular}[c]{@{}l@{}}The city of Auckland, New Zealand\\  is also known as Tāmaki Makaurau,\\ its original Māori name.\end{tabular}                                                                                                                                       \\ \hline
\textbf{\begin{tabular}[c]{@{}l@{}}Vernacular\\ names\end{tabular}}          & \begin{tabular}[c]{@{}l@{}}Place names that are informal or colloquial, \\ that can be alternative to formal, official or \\ administrative names, or might be present in \\ the absence of a formal name. Vernacular \\ names can sometimes be vague with \\ imprecise boundaries.\end{tabular} & \begin{tabular}[c]{@{}l@{}}Brum to refer to Birmingham in \\ England. Examples of vague \\ vernacular names are The South of \\ France, and Downtown Los Angeles.\end{tabular}                                                                                                  \\ \hline
\textbf{\begin{tabular}[c]{@{}l@{}}Temporal \\ changes\end{tabular}}         & \begin{tabular}[c]{@{}l@{}}Names of places change with time with \\ political,  administrative or socio-ethic \\ reasons.\end{tabular}                                                                                                                                                           & \begin{tabular}[c]{@{}l@{}}Istanbul used to be called \\ Byzantium and \\ Constantinople.\end{tabular}                                                                                                                                                                          \\ \hline
\textbf{\begin{tabular}[c]{@{}l@{}}Place name\\ ambiguity\end{tabular}}          & \begin{tabular}[c]{@{}l@{}}Two or more different places being called the same \\ place name.\end{tabular}                                                                                                                                                                                                & \begin{tabular}[c]{@{}l@{}}Lisbon is the capital of Portugal \\ but there are also several towns in \\ different states of the US called \\ Lisbon. The state of Wisconsin, USA \\ contains at least 4 inhabited palces \\ called Springfield.\end{tabular}                                                                                                                \\ \hline
\textbf{\begin{tabular}[c]{@{}l@{}}Entity \\ ambiguity\end{tabular}}         & \begin{tabular}[c]{@{}l@{}}A name could be shared by a geographic \\ place as well as another prominent person, \\ organization etc.\end{tabular}                                                                                                                                                & \begin{tabular}[c]{@{}l@{}}Liverpool could be referring to the \\ city in England or the Liverpool \\ football club.\end{tabular}                                                                                                                                               \\ \hline
\textbf{\begin{tabular}[c]{@{}l@{}}Region \\ specific \\ names\end{tabular}} & \begin{tabular}[c]{@{}l@{}}The same feature (especially large natural \\ features like mountains or rivers) may have \\ different names depending on the different \\ regions the feature crosses.\end{tabular}                                                                                  & \begin{tabular}[c]{@{}l@{}}The river Danube is called "Donau" \\ in Germany, "Dunaj" in Slovakia, \\ "Duna" in Hungary, "Dunav" in \\ Croatia, "Dunav" and "Дунав" in \\ Bulgaria, "Dunărea" in Romania and\\ in Moldova and "Dunaj" and "Дунай" \\ in the Ukraine.\end{tabular} \\ \hline
\textbf{Abbreviations}                                                       & Abbreviations of names.                                                                                                                                                                                                                                                                          & \begin{tabular}[c]{@{}l@{}}Los Angeles is abbreviated to L.A. or\\ S.F. for San Francisco.\end{tabular}                                                                                                                                                                         \\ \hline
\textbf{\begin{tabular}[c]{@{}l@{}}Name Sim-\\ -plifications\end{tabular}}   & \begin{tabular}[c]{@{}l@{}}Similar to abbreviations but some places have \\ simplifications for the name for reasons such \\ as the original name being harder to \\ pronounce or write.\end{tabular}                                                                                            & \begin{tabular}[c]{@{}l@{}}Place with the longest place name – a \\ place in New Zealand called “Taumat-\\ -awhakatangi­hangakoauauotamatea­tu-\\ -ripukakapikimaunga­horonukupokai-\\ when­uakitanatahu” is often simplified \\ as “Te Taumata”.\end{tabular}                  \\ \hline
\textbf{Nicknames}                                                           & \begin{tabular}[c]{@{}l@{}}Common nicknames are type of alternative \\ names. Similar to a vernacular name, but is \\ more likely to be used to embellish a \\ description of a place, rather than be commonly\\ used as an alternative to the official name.\end{tabular}                       & \begin{tabular}[c]{@{}l@{}}The capital of France, Paris, is also \\ known as the “City of light”.\end{tabular}                                                                                                                                                                  \\ \hline
\end{tabular}
\end{table}

The ADL gazetteer content standard (GCS) provided support to address some of the above issues, particularly with regard to enabling encoding of alternative versions of the name of geographic feature. It also specified a range of attributes of each place name and its variants \cite{hill1999indirect}. Listed below are some of them, but of these only time period was a required element (see also \cite{hill2006gazetteers}).

\begin{itemize}
    \item Official or authoritative source if any
    \item Etymology (derivation) of name
    \item Language code of the language the name was written in. (for example, ‘en’ for English; ‘fr’ for French; ‘mi’ for Māori)
    \item Pronunciation (text or audio file)
    \item Transliteration scheme
    \item Confidence in the name
    \item Abbreviations if any
    \item Time period. Three options – former, current and proposed
    \item Links/references to more information
\end{itemize}

Early digital gazetteers developed in the late 1900s compiled  places extracted from authoritative sources. These sources included  non-digital gazetteers, maps, and records from official toponymic authorities. In the first decade of this millennium the interest in web harvesting, and later the advent of VGI, meant gazetteers were no longer limited to authoritative sources, and hence could include non-official place names. This included vague place names and vernacular place names \cite{jones2008modelling}.

Vernacular place names are names that are commonly used regardless of whether they are official or not. They can also include vague place names that refer to locations lacking distinct boundaries, or places whose extents and landmarks might be perceived differently by various individuals and across different contexts \cite{davies2009user}. Their acquisition is vital for information retrieval tasks \cite{twaroch2008acquisition} in order to ensure such names are recognised when used in queries.  GumTree \cite{twaroch2008mining}, Craigslist \cite{hu2019natural}, phonebooks sites \cite{goldberg2009extracting}, social media sites like Flickr and Panoramio \cite{popescu2009mining} are examples of sources that have been harvested for vernacular names. Several methods employed in the extraction of vernacular place names from these websites can be identified throughout literature. One of the early methods was based on analysing content of scraped web documents using regular expressions to identify addresses and a list of spatial prepositions to detect place names, followed by a web search validation procedure \cite{twaroch2008mining}. They report about 20 place names obtained from GumTree, an advertising and community website, in and around Cardiff, UK, that were not found in the British Ordnance Survey OS50k gazetteer.

Vernacular names have also been extracted from web pages using targeted query methods. Thus queries could take forms such as “$\langle target \; region\rangle$”, “$\langle concept \rangle$ $\langle target \; region \rangle$” or “$\langle lexical \; pattern\rangle$ $\langle target \; region \rangle$” \cite{arampatzis2006web, jones2008modelling}. The target region is a name to be extracted, the concept can be a feature such as a hotel while the lexical pattern (also called a trigger phrase) could include spatial relations as “is [in \textbar \: located in \textbar \: situated in] the [center \textbar \: north \textbar \: south \textbar \: east \textbar \: west] of” where \textbar \: means or. NER methods can then be used to detect place names in the retrieved snippets of text.  An alternative, web-based approach to vernacular name detection is to mine the content of addresses on web pages, in which vernacular names of neighbourhoods can be found within the address structure, using regular expressions. The coordinates are then indicated by any accompanying postcode \cite{brindley2018generating}.

\cite{hu2019natural} use more sophisticated natural language processing techniques such as NER, i.e. the recognition of important nouns and proper nouns, on text scraped from Craigslist. They combine two off the shelf NER tools,  spaCy\footnote{https://spacy.io/}  and Stanford NER\footnote{https://nlp.stanford.edu/software/CRF-NER.shtml} , with two more case sensitive and Twitter re-trained Stanford variants \cite{hu2019natural}. They report their method can enhance gazetteers with new places (places that were not recorded in existing gazetteers)  especially for features like local neighbourhoods, parks, schools and points of interest along with other alternative vernacular names for places already reported in gazetteers such as GeoNames, TGN and WOF.

\cite{rice2012supporting} demonstrated the ability to rely fully on VGI for vernacular place names within a small area – the George Mason University. The gazetteer was constantly updated by students and staff with changes, repairs and closedowns within the university premises to help the visually impaired. A slightly different study was carried out to retrieve peoples’ perception of places and the vernacular names they use in \cite{twaroch2010web} where the authors created a web platform\footnote{https://peoplesplacenames.com/map.php} for people to name places within a familiar geographical location. This was done variously with an interactive map to name all the place names they knew within an area that they specified, using a postal code, or asking for names used in place of official names. While this may be an effective method to collect vernacular names, its extensibility depended largely on the interest of the crowd to participate in it. 

\subsection{Feature Type}
\label{sec:FeatureType}

The feature type is defined in \cite{hill2000core} as a type “selected from a type scheme of categories for places/features” \cite[p.1]{hill2000core}, and examples include natural features such as a river or a mountain, artificial features like a building, post office or village, and the administrative class or jurisdiction. Gazetteers vary according to whether only a single type is recorded for a place name instance, or whether multiple types are sometimes listed (as in the Thesaurus of Geographic Names). Here we discuss feature type schemes and the various methods used to store these schemes and how this choice has, at times, dictated the structure of the gazetteer as a whole.

\subsubsection{Feature Type Schemes}

Most digital gazetteers developed during the end of the 20\textsuperscript{th} century and the beginning of the 21\textsuperscript{}{st} century have developed their own feature type schemes \cite{hill2000core}. This situation has plagued digital gazetteers and left a persistent problem in gazetteer interoperability, leaving even modern gazetteer developers and users grappling with the issue of inconsistency in the use of descriptors or codes for feature type. For example, in GNIS, which uses a broad categorization binning many discrete types together, “Park” could refer to reserves, forests, monuments, lakes, historical or archaeological sites and even cemeteries and museums. On the other hand a “Park” in a scheme that is more specific, like GeoNames, means an often forested place maintained for recreation and beauty. GeoNames gives cemeteries, reserves, forests, lakes, and historical sites their own codes. This problem dates back to the beginning of authoritative digital gazetteers in an era where digitization was still new, and perhaps interoperability was not such a high priority. The UN ECOSOC\footnote{https://ecosoc.un.org/en} attempted to build a standard for gazetteers, and was mostly focused on global coverage. It was not adopted in the numerous gazetteers developed by individual governments that used their own sets of feature types, reflecting different cultures, languages and countries \cite{rsingh2016InternationalStandards}.

While online gazetteers, such as those of the U.S Geological Survey and the U.S. National Geospatial-Intelligence Agency (formerly National Imagery and Mapping Agency), Australian Geographic Names Gazetteer and the Getty Thesaurus of Geographic Names, used their own feature type schemes, the ADL FTT (Feature Type Thesaurus) was the first large scale effort to merge several of these schemes. The FTT uses types found in the schemes of its source gazetteers and place name records, which are referred to as lead-in terms that are linked to their corresponding preferred terms in the FTT \cite{hill2000core, hill2006gazetteers}.  While manually mapping the existing schemes to their own, the ADL FTT also generates feature types using common “surnames” of places. “Surnames” refer to commonly found end positions of place names (e.g. the last word of the place name Waikato River indicates its feature type). They captured not only single word feature types that are commonly found after proper nouns of place names but also multiple word feature types like “oil seep”, “gas field” etc \cite{hill2000core}. The ADL FTT included 209 preferred terms that accommodated 978 lead-in terms from existing feature type schemes. The FTT consisted of 6 top level feature types: administrative areas, hydrographic features, manmade features, physiographic features, land parcels, and regions. These top level features were further broken into 3 more levels.

The ADL data model along with the FTT became adopted by numerous gazetteers \cite{reid2003geoxwalk, vestavik2007merging, manguinhas2008geo, liu2009kidgs}. There were also attempts to enrich the feature types in the FTT using further extraction of generic terms, correlating them to types already in the FTT and placing them in suitable positions in the FTT. Thus \cite{wang2006automatic} used commonly occurring tokens in place names that show correlations with their respective feature type as potential lead-in terms to the FTT, using hierarchical clustering to identify the correct insertion location of the FTT. The authors of \cite{wang2006automatic} note that their efforts led to the identification of numerous non-English generic terms. Other gazetteers that have global coverage, notably GeoNames, have their own feature type schemes, which were in turn sometimes adopted by other gazetteers \cite{popescu2008gazetiki}.

Other efforts by gazetteer developers to create their own type schemes included The Information Commons Gazetteer with a simple scheme that accommodated places from GNIS and the US National Geo-spatial Agency’s GEOnet Names Server (GNS) database \cite{lucas2006information}. Adherence to a unified data model or feature type scheme was challenging for several reasons:

\begin{itemize}
    \item Difference in use case of gazetteers - requiring different feature types classification schemes. For example, a general purpose gazetteer might include all historical or archaeological sites under a general type but this is inadequate for a historical gazetteer that could require specialised categorization of the various historical sites.
    \item Disagreement on preferred names – e.g.  gazetteer developers may choose to use synonyms of the terms used in the source gazetteer thesauri. Thus the ADL FTT lists lead-in terms (from its sources) that are synonymous to the terms they have chosen to represent a concept.
    \item Cultural biases – The descriptions and the vocabularies used in different gazetteers are culturally biased and may not suit all users from all regions and cultures. The meaning of a place referred to by a name can vary between cultures, and hence the naming of places varies with the prominent language of these regions. Another issue is the under representation of some languages, cultures and regions due to historical events and power structures.

\end{itemize}

\subsubsection{Feature Type Scheme Formats}

Here we discuss feature type schemes in gazetteers and their relative merits, with a focus on thesauri and ontologies. Thesauri were used in some early digital gazetteers, notably the Getty Thesaurus of Geographic Names. However, their use is accompanied by some shortcomings. In particular:

\begin{itemize}
    \item Thesauri only supported a limited number of relations, typically of hierarchical, associative and equivalence \cite{obrst2010ontologySpectrum}.
    \item Most feature type thesauri did not enable multiple hierarchies, making it difficult to represent some entities as perceived by humans.
    \item Entities in thesauri may not have specific characteristics that govern the establishment of relations,  as entities are not as robustly defined as concepts in ontologies, in which definitions of concepts can govern their relations. 
    \item The definition of a concept as an entry in a thesaurus is informal and may be insufficient for further integration and expansion in different applications.
    \item Problems in interoperability such as resulting from the previous point.
\end{itemize}

It is to overcome these issues that gazetteer developers started adopting alternative feature type approaches, particularly ontologies. Interest in using ontologies (geo-ontologies) for feature types in gazetteers coincided with the emergence of the concepts of the semantic web and linked data \cite{lassila2001semantic, bernerslee2006linkeddata, bizer2011linked}. The more formal nature of an ontology and the fact that it can be regarded as a shared concept among various parties \cite{borst1997constructionontologies} can be seen to help alleviate some of the issues with thesauri. A comparison between the use of thesauri and ontologies for feature type schemes is shown in Table 2.

\begin{table}[]
\small
\caption{Thesaurus and ontology comparison}
\begin{tabular}{l|ll}
\hline
                                                                                    & \multicolumn{1}{c}{\textbf{Thesaurus}} & \multicolumn{1}{c}{\textbf{Ontology}} \\ \hline
\textbf{Application}                                                                & Information retrieval and structuring.                                                  & \begin{tabular}[c]{@{}l@{}}Inference and reasoning, information  \\retrieval.\end{tabular}                                                                              \\
\textbf{Enforced hierarchy}                                                         & \begin{tabular}[c]{@{}l@{}}Generic, whole-part or instance \\hierarchy. \end{tabular} & is-a hierarchy. \\                                                                                                                                                         \\
\textbf{Relations supported}                                                        & Restricted number of relations.                                                         & Arbitrary number of relations.  \\                                                                                                                                        \\
\textbf{Formal semantics}                                                           & \begin{tabular}[c]{@{}l@{}}No formal semantics to support reas-\\ -oning.\end{tabular}  & \begin{tabular}[c]{@{}l@{}}Formal semantics enables deductive \\ reasoning. This allows inference of new \\ knowledge about relations between \\ concepts.\end{tabular} \\
\textbf{\begin{tabular}[c]{@{}l@{}}Definition of terms\\ and concepts\end{tabular}} & Terms representing concepts.                                                            & Formal specifications of concepts.                                                                                                                                      \\ \hline
\end{tabular}
\end{table}

Early attempts to build ontologies for gazetteers were based on existing gazetteer feature type schemes \cite{janowicz2008role}. With ontological approaches, the boundary between an ontology as a feature type scheme and building a place name ontology that could be populated with instances of places and used as a gazetteer begins to blur \cite{ping2009building, southall2009organisation, machado2010ontological, grenoble2019ontology}.

Adopting an ontology based approach for a gazetteer claims several advantages over the traditional list/taxonomy or thesaurus based approaches. In \cite{fu2005ontology, delboni2007semantic}, the authors point out that an ontology based approach can improve spatial querying with regards to vague place names such as “South of France”. The ability to define arbitrary relations among concepts that can aid in querying, and reasoning is another advantage of ontologies. Relations like \textit{hasOrigin} or \textit{hasDestination} are examples in which a stream can have a spring as an origin, and a tributary can have a river, or a river an ocean, as its destination. \cite{peng2010folksonomy, souza2005role} present an ontological gazetteer that explicitly stores topological and vague relations such as “within”, “touch”, and “near”. GeoSPARQL \cite{car2022geosparql, battle2012enabling}, which is an extension to SPARQL \cite{hommeaux2008sparql}, by the Open Geospatial Consortium (OGC)\footnote{https://www.ogc.org/}, allows querying semantic geographic data on the web using 9-IM/RCC topological relations \cite{clementini1994modelling, clementini2014rcc}, further improving the functionality of ontology-based approaches \cite{laurini2015geographic}. Improved reasoning in Geographic Information Retrieval (GIR) applications is made possible due to the expressivity of description logic associated with some ontologies – thus enabling subsumption and similarity based reasoning \cite{janowicz2008role}. Further benefits of ontologies are that they can support multiple hierarchies and have the potential to be more maintainable, and scalable and to facilitate interoperability. The benefits are however offset by higher costs of implementation.

It is important to note that while some of the most widely used gazetteers today, like GeoNames, Nominatim and the Getty Thesaurus of Geographic Names may have published their data as Linked Data and built ontologies around them, they are still operating primarily with either simple feature type taxonomies or feature type thesauri.

\subsection{Geographic Footprint}

The geographic footprint is the quantitative geometric entity that links a toponym to its real world location. Latitude and longitude coordinate pairs in the form of points, bounding boxes, lines, polygons and grid references can be used to represent the geographic footprint of a place. For instance, in the ADL Gazetteer Content Standard (GCS), it is compulsory for each feature to have one or more geometries (points, lines or polygons) and a bounding box which is a generalisation of the geometries. It is possible for a single place to have multiple geometric representations -
\begin{itemize}
    \item from different sources that represent the same place differently
    \item different footprints representing the changes of the location over time
    \item represented with multiple types – for example a point and a bounding box, or a point and a polygon which represent the object at different levels of generalisation 
    \item multiple representations of linear or polygon geometry according to level of generalisation
\end{itemize}

Geometric footprints stored in gazetteers are clearly very much an approximation of the location of the represented place. Spatially extensive features such as cities, lakes, or rivers are commonly represented just by a zero-dimensional point, but almost all real world places require at least two dimensional representation for better fidelity. Lines are sometimes used to represent linear features like roads or rivers, although all such natural linear features have a width. Bounding boxes can provide an indication of extent of a feature but how well they do so varies with its actual form. Thus, for some features like cities it could be quite useful but for some linear features it might grossly overestimate the actual area extent. 

While a polygon with a possibly large number of vertices can be used to model the boundary of some real world features with high accuracy, natural features do not usually have crisp boundaries and therefore their representations are still inevitably approximations to some degree. Very detailed geometry can also introduce computational overheads. Due to the unavoidable imprecisions introduced into gazetteers when modelling locations, \cite{hill2000core} has suggested that gazetteers should convey the degree of approximation.

\subsubsection{Approximating the geometric extent of gazetteer features}

In \cite{wilson2004new}, the authors propose using circles for footprints when attempting to derive the extent of populated places with only point locations. They use the mean distance between point pairs within a particular region of a county to calculate a radius for these point buffers. Although intended to represent the fuzziness in the boundaries of populated places the method may have varying suitability depending on the density of named places and their population density, and is clearly focused on non-natural features. \cite{singh2018strategies} focus on improving minimum bounding boxes (MBB). They test out hierarchical, geometrical, probabilistic and heuristic methods to represent a Minimum Bounding Rectangle (MBR) and test their methods on data from GeoNames and OSMnames\footnote{https://osmnames.org/}  using Google map MBBs to verify. They argue their probabilistic approach can be used to improve footprints of gazetteer place names at district level or larger while their heuristic and geometrical approaches can be used to improve data on places that do not have enough coverage. They use the parent-child relationships given in gazetteers to model the MBR of the parent place and use optimization methods to include as many children as possible but also to reduce the distance between the point location given for the parent place and the center of the MBR they derive.

Given prior knowledge of the containment and non-containment of point-referenced places within larger regions, that could be vague vernacular places, Voronoi diagram methods were applied in \cite{alani2001voronoi}. In an approach to modelling the extent of sets of points contained within vague regions with known names, obtained through web search methods, several methods based on Delaunay triangulation were applied in \cite{arampatzis2006web}. The boundaries of vague place names were estimated in \cite{zhang2010contextual} based on their natural language spatial relationships to places with known coordinates from Chinese gazetteers. They generate boundaries by combining geometric applicability models (also known as spatial templates) of multiple individual natural language spatial relationships to the known places.

\subsubsection{Modelling extents with kernel density estimation (KDE)}

One of the most common methods for representing the extent of places, based on multiple point data samples for a named place retrieved from VGI and web pages, is to apply kernel density estimation (KDE) to the points \cite{hu2019natural, twaroch2008acquisition, brindley2018generating, thurstain2000defining, hollenstein2010exploring, twaroch2019investigating, li2012constructing}. When using social media, a significant problem can arise with introducing bias in the inferred region \cite{hollenstein2010exploring}. Thus, individual contributors could skew the data to a particular person’s perception of the location of a place. Hence when there are multiple contributors to a set of data it could be appropriate to limit the number of contributions from each user. Bulk uploads can also skew results, thus if there are many tags with identical coordinates only one such coordinate might be selected. A problem specific to KDE is deciding on the boundary of the resulting fuzzy region. Various threshold values for the volume of the KDE surface have been proposed, such as between 50\% and 90\%, e.g.  \cite{jones2008modelling, brindley2018generating, hollenstein2010exploring}. Similarly different KDE bandwidth values have been used, in the case of \cite{brindley2018generating} there being two, for different levels of granularity, each in combination with thresholds. As an alternative to KDE for modelling sets of points, fuzzy set based methods with thresholds were applied in \cite{gao2017constructing}.

\subsubsection{Discrete global grids}

Discrete global grid systems are another technique used for grounding place names (i.e. generating geometric representations for them) in gazetteers, though not many were identified in our study. Wāhi \cite{adams2017wahi} is a discrete global grid gazetteer that implements three grid reference systems.

\begin{figure}[H]
\centering 
\includegraphics[width=0.8\textwidth]{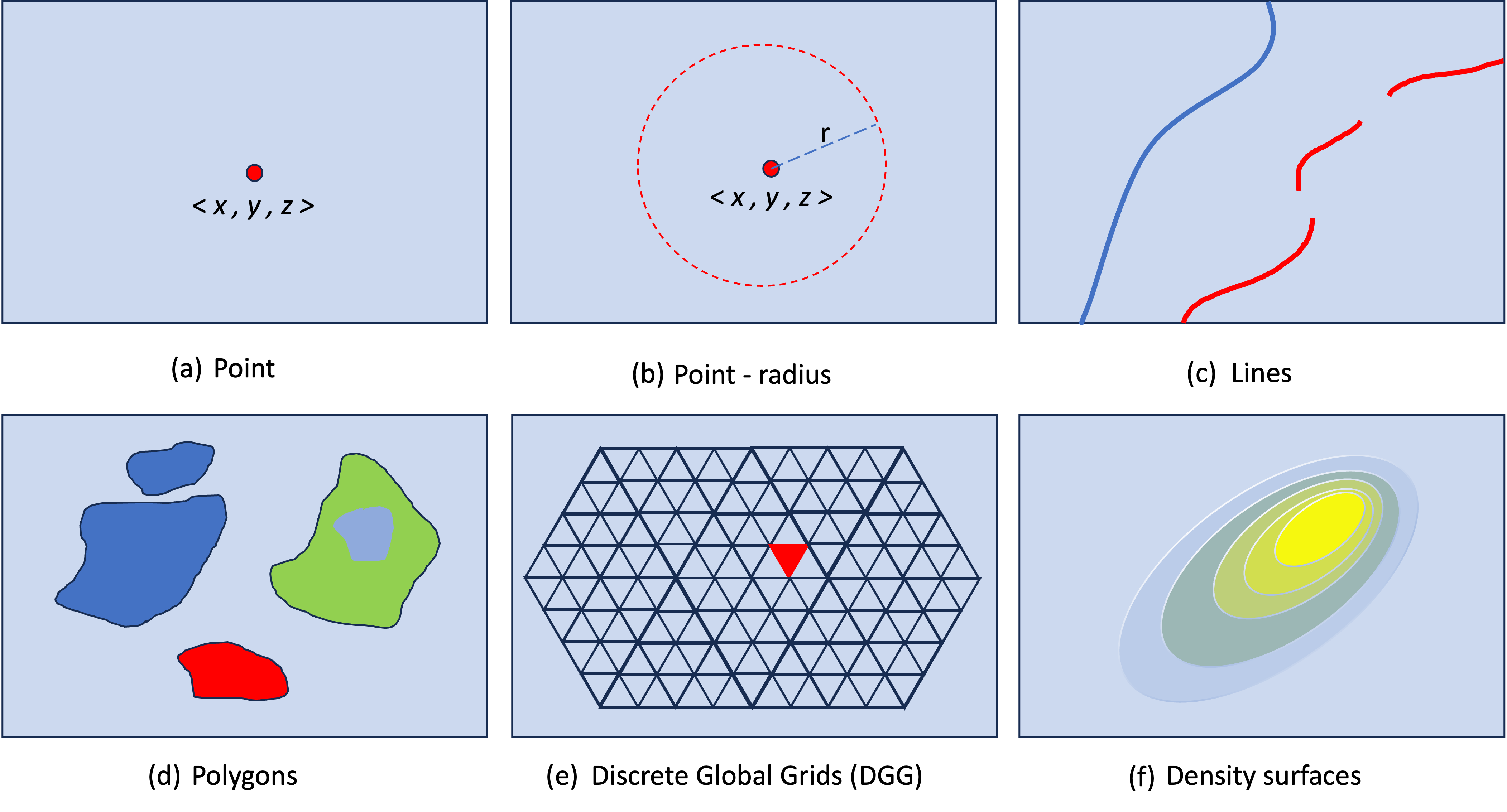}
    \caption[The system.]{Various methods of recording geographic footprints. As shown in (a), a point is defined by an 'x' and a 'y' coordinate pair. Gazetteers may also optionally store a third coordinate for the vertical height or elevation from some reference level. (b) shows a point-radius representation where a radius distance, 'r', accompanies the point. Lines can be represented as a continuous collection of points or a collection of disjoint line segments as shown in (c). (d) shows different types of polygons: a simple polygon in red, a multi-polygon in blue and a complex polygon in green which has a hole in the center. (e) shows a location in a discrete grid in red. Density surfaces produce a raster grid with each cell representing a density as shown in (f).}
\label{Fig4}
\end{figure}

Fig. \ref{Fig4} illustrates some of these methods used to record geographic footprints.

\subsubsection{Temporal change in footprints}

Some research has focused on the update of footprints to reflect the gradual change with time of features such as human settlements and forests, lakes and rivers. Satellite images \cite{newsam2008integrating} and aerial images \cite{agouris2000capturing} of these features can be used to enhance gazetteers with changes over time. \cite{newsam2008integrating} used satellite images with visual appearance models to estimate the bounding boxes, while \cite{agouris2000capturing} use a least squares matching algorithm to detect changes in boundaries of features, comparing the boundary stored in their spatio-temporal gazetteer with new images of the feature. With the abundant availability of satellite imagery datasets and advances in computer vision, automatic generation and temporal maintenance of complex footprints from these sources has considerable potential.

\subsubsection{Efficiency in footprint storage}

The alternative ways of representing footprints alluded to earlier are accompanied by issues of storage and computational efficiency. Not all applications need features to be delineated in high detail.  In this context \cite{hill2000core} introduces satisficing into digital gazetteer footprints, i.e. not to seek optimal solutions to certain problems because the costs are too high, but to settle for solutions that are satisfactory given the cost. With the advent of open and hence sharable linked data, the requirement of many gazetteers to store complex geographic objects comes into question, as opposed to referencing a particular geometric representation. Thus storage of duplicate complex geometries can be avoided \cite{regalia2018gnis, lonneville2021publishing}, though it does not necessarily reduce the computational overhead for the end user.

\subsection{Temporality}

\cite{hill2006gazetteers} identifies the need to record temporality in every major component of the gazetteer and hence achieve a truly spatio-temporal gazetteer. In the GCS, every place name, feature type, footprint, relationship, classification and other item of data associated with the place (including attributes such as population) needs to have a time period associated with it. In addition to start and end time data items, in association with labels of current, former or proposed, the GCS supports recording a detailed time period, a named time period and notes on the time period.  

Temporal data can reflect change of urban boundaries, civilizations, kingdoms, rulers and governing states as well as aspects of the continuous change in the natural environment. Changes in the name of a place might affect our perception of it. It could be argued that changes in the name of city due to change in the governing regime but not to a large extent in the population, could be perceived differently from, say, the change in a school from a secular to a religious one in which the entire ethos and much of the student population changed. An example of the former is Saint Petersburg in Russia being Petrograd in the early 20th century, before becoming Leningrad in honour of the Russian Communist leader, and then reverting to Saint Petersburg with the fall of communism. The perception might change according to the prior knowledge or experience of the perceiving individual, but also in the nature of the change. In the case of the school it might be regarded as less different if only the name changed but everything else remained the same. Similar issues could arise in the change in the feature type of a place. The change of geometric footprints could also result in different perceptions, for example if a small settlement evolves to a large conurbation or a lake dries up or grows, with respective environmental impacts.  

Temporality becomes a key focus when dealing with historical gazetteers. In \cite{southall2009organisation}, the authors point out the difficulty in recording or finding out when a place started or stopped being called by a name. They point out that even vague dates are hard to find when searching for information and argue that gazetteers should record as many historical names as possible and treat the usage of names as attestations of their validity. 

Capturing temporal changes using aerial images \cite{agouris2000capturing} and satellite images \cite{newsam2008integrating} was discussed earlier under geographical footprints. In order to be temporally scoped it is not only important to update the footprints but to timestamp these footprints.

Following the ADL GCS, many gazetteer developers accommodated some sort of temporal attributes to their models \cite{cardoso2014gazetteer, nagata2019community, ducatteeuw2021developing, de2021digital}. Furthermore, the latest formats for storing place records, especially on the linked web, facilitate temporal scoping. The GeoJSON-T format is an enhancement to the GeoJSON format (the T stands for temporal) and the latest Linked places format is built on top of the GeoJSON-T format \cite{ducatteeuw2021developing}.

\subsection{Stored and Inferred Relations Between Places}

One aspect of the place name component is that a name may include or be directly associated with its hierarchical parents, particularly administrative regions, such as county, state, province and nation. In addition to such specified hierarchical information, some state agencies and research projects developed gazetteers that also supported functionality to compute topological relations from footprint geometry \cite{tochtermann1997using, riekert2002automated, weaver2003digital}. Related place name knowledge resources referred to variously as ontological gazetteers \cite{machado2010ontological} and place name ontologies \cite{jones2001geographical, ping2009building} stored spatial relations explicitly rather than computing on the fly. In the case of \cite{jones2001geographical} the model included reference to application-specific data items (artefacts) as well as a feature types from an external feature type thesaurus.

\subsection{Other}
While the main components of gazetteers are those described above, some gazetteers do include other types of information that can go some way towards the vision of the richer representation of the named place that was referred to in the introduction. For example, the TGN has a category of Place Type, which while equivalent in some ways to a feature type (and includes the main feature type) is an explicit recognition of the concept of place. Its entry for the Italian city of Venice is as follows:

\begin{verbatim}

inhabited place (preferred, C)	 .... settled by refugees from Lombard invasion
                                      after 568
city (C)	
commune (administrative) (C)	
regional capital (C)	
provincial capital (C)	
cultural center (C)	 ............ 	flourished during 15th-18th cen.; home to many 
                                    important Renaissance artists & architects
trade center (C)	
military center (H)	
republic (H)	 ................ 	from middle of 12th cen. until 1797
World Heritage Site (C)	 ......... 	since 1987
first level subdivision (C)	

\end{verbatim}

The ADL GCS originally included a single "description" field to store a free text description of a place, but this was later modified to allow multiple descriptions reflecting the different facets of a place. The Pleiades gazetteer described as “A Gazetteer of Past Places”\footnote{ https://pleiades.stoa.org/places} also records place types which typically refer to cultural origins, such as “sanctuary (religious center)”. This gazetteer also often includes free text descriptions of the place referring to its history.
Examples of other types of content include “History” and “Description” fields in the GNIS, where the latter is often used to describe the geographic context of a place. The New Zealand NZGB gazetteer\footnote{https://gazetteer.linz.govt.nz/} has  “Events” and “History/origin/meaning”, and the TGN has a “Notes” field usually used to describe the historical origins as well as references to important cultural sites. The TGN Notes entry for Venice reads as follows: 

\emph{The islands of the Venetian lagoon were settled by refugees from Altino and Aquileia after the Lombard invasion of northern Italy in 568. The settlement expanded into the Adriatic and Aegean from the 11th century. It was ruled by Austria, then France, 1798-1848. It went to Italy in 1866 The entire city including the lagoon was named an UNESCO World Heritage Site in 1987.}

Some of the richest place related content is to be found in resources that are not described as gazetteers. One of the most widely used examples is the set of the place specific pages in Wikipedia, which are complemented by the more structured representations of DBpedia and Wikidata which include data properties such as \emph{inception} (a date), \emph{significant event}, \emph{birth place} of, \emph{death place} of, \emph{comment}, \emph{abstract} (a summary of the Wikipedia content), \emph{residence} of, \emph{garrison} of, \emph{type} (including Yago\footnote{https://yago-knowledge.org/} place types). Other example resources are more localised as in Kā Huru Manu (The Ngāi Tahu Cultural Mapping Project)\footnote{https://kahurumanu.co.nz} from New Zealand which focuses on the Māori origins and stories associated with named places.

\section{Gazetteer Sources and Integration}
\label{sec:integration}

In this section, we explore the sources used to compile modern gazetteers and methods used in their integration. Most gazetteers and papers that were retrieved in our search, produced in the late 20\textsuperscript{th} century or 21\textsuperscript{st} century, were compiled by integrating other gazetteers. All modern digital gazetteers retrieved from our search fall into one of the following types depending on their sources:

\begin{enumerate}
    \item Gazetteers formed by the digitization of (and integration of one or more) historical gazetteers.
    \item Gazetteers formed by toponym extraction from existing maps either modern or historical.
    \item Gazetteers formed by integrating other digital gazetteers.
    \item Gazetteers formed by integrating VGI (or web harvested information) with existing gazetteers.
    \item Gazetteers formed completely from VGI or web harvested data.
\end{enumerate}

Table 3 shows the distribution of research papers in our review over time that used each of these sources, and indicates that most of the recent gazetteers were created by integrating or improving one or more already existing authoritative or VGI digital gazetteers.

\begin{table}[]
\caption{Distribution of various sources that have been integrated to compile gazetteers over the years. Each number refers to a citation.}
\begin{tabular}{rclllll}
\cline{2-7}
\multicolumn{1}{l|}{}                                                                                                & \multicolumn{1}{l|}{}                                                    & \multicolumn{1}{l|}{}                                                     & \multicolumn{1}{l|}{}                                                                                & \multicolumn{1}{l|}{}                                                             & \multicolumn{1}{l|}{}                                                                            & \multicolumn{1}{l|}{}                                                                  \\
\multicolumn{1}{r|}{\begin{tabular}[c]{@{}r@{}}Integrating \\ authoritative \\ gazetteers\end{tabular}}              & \multicolumn{1}{c|}{\begin{tabular}[c]{@{}c@{}}\cite{hill1999indirect},   \cite{cervellati2000composite},\\ \cite{mceathron2001naming}\end{tabular}} & \multicolumn{1}{c|}{\begin{tabular}[c]{@{}c@{}}\cite{reid2003geoxwalk}, \cite{lucas2006information}\end{tabular}} & \multicolumn{1}{c|}{\begin{tabular}[c]{@{}c@{}}\cite{brauner2007towards}, \cite{vestavik2007merging},\\ \cite{hung2008construction}, \cite{hastings2008automated}, \\ \cite{yang2010spatio}\end{tabular}}                        & \multicolumn{1}{c|}{\cite{yun2013efficient}}                                                           & \multicolumn{1}{c|}{\begin{tabular}[c]{@{}c@{}}\cite{pai2015gazetteer}, \cite{berman2016historicalgazsystem},\\ \cite{adams2017wahi}, \cite{hara2017digital},\\ \cite{hara2018digital}\end{tabular}}              & \multicolumn{1}{c|}{\begin{tabular}[c]{@{}c@{}}\cite{grossner2021linked}, \cite{lonneville2021publishing},\\ \cite{polczynski2022lessons}\end{tabular}}           \\
\multicolumn{1}{l|}{}                                                                                                & \multicolumn{1}{l|}{}                                                    & \multicolumn{1}{l|}{}                                                     & \multicolumn{1}{l|}{}                                                                                & \multicolumn{1}{l|}{}                                                             & \multicolumn{1}{l|}{}                                                                            & \multicolumn{1}{l|}{}                                                                  \\
\multicolumn{1}{r|}{\begin{tabular}[c]{@{}r@{}}Digitizing ancient \\ gazetteers/ \\ collection records\end{tabular}} & \multicolumn{1}{c|}{}                                                    & \multicolumn{1}{c|}{\cite{genoways2003results}, \cite{kelley2006resolving}}                                               & \multicolumn{1}{c|}{\cite{southall2009organisation}}                                                                              & \multicolumn{1}{c|}{\cite{yongtao2012literati}, \cite{cardoso2014gazetteer}}                                                       & \multicolumn{1}{c|}{\begin{tabular}[c]{@{}c@{}}\cite{blank2015geocoding}, \cite{goodchild2016gazpresentspatial},\\ \cite{pai2015gazetteer}, \cite{bekele2016spatiotemporal},\\ \cite{cardoso2016swi}, \cite{hara2017digital},\\ \cite{widgren2018mapping}, \cite{kirk2018assimilating}\end{tabular}} & \multicolumn{1}{c|}{\begin{tabular}[c]{@{}c@{}}\cite{chen2020local},\cite{ducatteeuw2021developing}, \\ \cite{de2021digital},\cite{polczynski2022lessons}, \\ \cite{liu2022making}\end{tabular}} \\
\multicolumn{1}{l|}{}                                                                                                & \multicolumn{1}{l|}{}                                                    & \multicolumn{1}{l|}{}                                                     & \multicolumn{1}{l|}{}                                                                                & \multicolumn{1}{l|}{}                                                             & \multicolumn{1}{l|}{}                                                                            & \multicolumn{1}{l|}{}                                                                  \\
\multicolumn{1}{r|}{Digitizing maps}                                                                                 & \multicolumn{1}{c|}{}                                                    & \multicolumn{1}{c|}{}                                                     & \multicolumn{1}{c|}{\cite{mostern2008historical}}                                                                              & \multicolumn{1}{c|}{}                                                             & \multicolumn{1}{c|}{\begin{tabular}[c]{@{}c@{}}\cite{hara2017digital}, \cite{yoshikatsu2017geographic},\\ \cite{li2018intelligent}\end{tabular}}                        & \multicolumn{1}{c|}{\begin{tabular}[c]{@{}c@{}}\cite{yoshikatsu2017geographic}, \cite{hara2017digital},\\ \cite{chen2020local}, \cite{horne2020beyond},\\ \cite{lin2020displaying}\end{tabular}}   \\
\multicolumn{1}{l|}{}                                                                                                & \multicolumn{1}{l|}{}                                                    & \multicolumn{1}{l|}{}                                                     & \multicolumn{1}{l|}{}                                                                                & \multicolumn{1}{l|}{}                                                             & \multicolumn{1}{l|}{}                                                                            & \multicolumn{1}{l|}{}                                                                  \\
\multicolumn{1}{r|}{\begin{tabular}[c]{@{}r@{}}Integrating VGI \\ and Authoritative \\ sources\end{tabular}}         & \multicolumn{1}{c|}{}                                                    & \multicolumn{1}{c|}{}                                                     & \multicolumn{1}{c|}{\begin{tabular}[c]{@{}c@{}}\cite{twaroch2008acquisition}, \cite{twaroch2008mining},\\ \cite{manguinhas2008geo}, \cite{popescu2009mining},\\ \cite{smart2010multi}\end{tabular}}                 & \multicolumn{1}{c|}{\begin{tabular}[c]{@{}c@{}}\cite{beard2012semantic}, \cite{yoshioka2013construction},\\ \cite{tanasescu2014reverse}, \cite{moura2014integration}\end{tabular}} & \multicolumn{1}{c|}{\cite{pradeepa2016construction}, \cite{regalia2018gnis}}                                                                     & \multicolumn{1}{c|}{}                                                                  \\
\multicolumn{1}{l|}{}                                                                                                & \multicolumn{1}{l|}{}                                                    & \multicolumn{1}{l|}{}                                                     & \multicolumn{1}{l|}{}                                                                                & \multicolumn{1}{l|}{}                                                             & \multicolumn{1}{l|}{}                                                                            & \multicolumn{1}{l|}{}                                                                  \\
\multicolumn{1}{r|}{VGI sources}                                                                                     & \multicolumn{1}{c|}{}                                                    & \multicolumn{1}{c|}{\cite{zhou2004discovering}}                                                  & \multicolumn{1}{c|}{\begin{tabular}[c]{@{}c@{}}\cite{rattenbury2007towards}, \cite{ahern2007world},\\ \cite{jones2008modelling}, \cite{popescu2008gazetiki},\\ \cite{goldberg2009extracting}, \cite{popescu2009mining},\\ \cite{kessler2009bottom}, \cite{peng2010folksonomy}\end{tabular}} & \multicolumn{1}{c|}{\begin{tabular}[c]{@{}c@{}}\cite{bosca2009automatic}, \cite{rice2012supporting},\\ \cite{gelernter2013automatic}, \cite{lamprianidis2014extraction}\end{tabular}}  & \multicolumn{1}{c|}{\begin{tabular}[c]{@{}c@{}}\cite{de2015leveraging}, \cite{pradeepa2016construction},\\ \cite{oliveira2016gazetteer}, \cite{gao2017constructing},\\ \cite{mcdonough2017mapping}\end{tabular}}         & \multicolumn{1}{c|}{\cite{hu2019natural}}                                                                \\
\multicolumn{1}{l|}{}                                                                                                & \multicolumn{1}{l|}{}                                                    & \multicolumn{1}{l|}{}                                                     & \multicolumn{1}{l|}{}                                                                                & \multicolumn{1}{l|}{}                                                             & \multicolumn{1}{l|}{}                                                                            & \multicolumn{1}{l|}{}                                                                  \\ \cline{2-7} 
                                                                                                                     & 1999-2002                                                                & \multicolumn{1}{c}{2003-2006}                                             & \multicolumn{1}{c}{2007-2010}                                                                        & \multicolumn{1}{c}{2011-2014}                                                     & \multicolumn{1}{c}{2015-2018}                                                                    & \multicolumn{1}{c}{2019-2022}                                                          \\
                                                                                                                     & \multicolumn{6}{c}{Publication Date}                                                                                                                                                                                                                                                                                                                                                                                                                                                                                                       
\end{tabular}
\end{table}

As evident throughout the literature, all the methods listed above excluding (2) and (5) have gazetteer  integration as a key element in gazetteer compilation. Gazetteer integration in general requires identification of duplicates between entries in the gazetteers and then resolution of differences to combine them as a single record. Both these tasks are complicated by the existence of multiple names for the same place and different feature type schemas (potentially resulting in different feature types for the same named place), and the variety of geographic footprints available along with their source inaccuracies.

Duplicate identification or entity resolution, also known as gazetteer matching in this context, is the process of matching features (gazetteer records) that refer to the same real world place, either between gazetteers or within the same gazetteer. Methods for resolution of duplicates often use a combination of string similarity between toponyms, geographical distances between the features  (or other forms of geometric relationship like overlap) \cite{sehgal2006entity, zheng2010detecting} and a semantic distance based on the feature types. In this section we summarise some of the methods used in the literature. Note that nearly all of the cited papers refers to gazetteers. There is however a body of literature that applies similar methods to match points of interest (POI). We have cited a couple of recent such papers as they employ techniques that are relevant to gazetteers \cite{isaj2019multi, balsebre2022geospatial}.

The duplicate identification task can be classed as either intra-gazetteer or inter-gazetteer. Inter-gazetteer matching is considerably more complex due to the challenges in matching different feature type schemas \cite{hill1999indirect}. Appendix Table 1 presents a summary of publications that either present methods for gazetteer integration or present gazetteers that have been compiled by integrating other gazetteers and that discuss deduplication techniques. This table has columns indicating whether integration is inter- or intra-gazetteer or both; the data sources; the main types of information used for matching records with regard to the use of the names, footprints and feature types; and notes on how the features are exploited focusing on the types of classifier or other statistical methods employed in machine learning techniques. In column Feature Name we list the techniques for name matching applied in cited papers. These mostly refer to string similarity measures such as Levenshtein edit distance, and Jaro-Winker, Jaccard, Monge-Elkan, and Soundex similarity and distance measures. Geographic footprint similarity measures are based on distance between the footprints, the degree of overlap, or containment within other footprints. The Feature Type column refers to the use of explicit feature categories and any text that might indicate aspects of the feature type (e.g. ‘eating’). Here similarity measures include distance between terms in a class hierarchy, distance to a common parent in a class hierarchy, presence or absence of the type in a vector of all types (i.e. one-hot encoding), as well as Dice coefficient, and the Jaccard and the Wu and Palmer distance between the type terms. Feature type terms, and feature names, can also be represented as word embeddings when input to machine learning classifiers.

The various measures of similarity have either been used with explicit thresholds in rule-based methods, or as input features for a machine learning classifier. In the case of more recent deep learning transformer-based methods \cite{balsebre2022geospatial} similarity is learnt from the word embeddings in the training data, without use of explicit string similarity metrics.

Most of the papers listed in Appendix Table 1 refer to applications of integration techniques that were applied when building a gazetteer, but these techniques could also be applied in a more dynamic context in which a gazetteer service provides online access to multiple external sources that are matched and integrated at the time of user query. This approach has been referred to as a meta-gazetteer \cite{smart2010multi} and has similarities with for example the extraction, transformation and loading (ETL) approach of \cite{manguinhas2008geo}.

Except for the techniques in \cite{moura2014integration,manguinhas2008geo, cheng2010data}, the methods summarized in Appendix Table 1 are applicable for instance level matching, independent of the underlying KOS. However, schema level gazetteer alignment or the alignment of the underlying data structures is a broader research topic. The alignment of two gazetteer KOS creates a new KOS with a new typology of features that match and retain the typology or concepts of the original sources to varying extents. Generic ontology alignment is well researched \cite{shvaiko2005survey, ardjani2015ontology}. However, standards and frameworks for place ontology alignment are less common and rarely replicated. 

Generically, ontology alignment can be broadly categorized as concept level alignment or structure level alignment \cite{shvaiko2005survey}. During alignment, a concept or element level alignment compares labels and definitions of concepts or terms in ontologies or thesauri respectively. This approach aims to exploit lexical or semantic similarities of the  concepts. Conversely, structure level alignment considers concepts’ relations to other concepts and their relative placement within a KOS (e.g., within a graph or a tree structure). We note that most of the gazetteer thesauri or geospatial ontology alignment techniques are dominated by concept level alignment approaches. Therefore, it is more apt to categorize the gazetteer ontology and thesauri alignment depending on whether the methods used are manual or automatic (see below). 

Gazetteer data structures are manually aligned often during the creation of a new gazetteer \cite{cheng2010data, manguinhas2008geo, grenoble2019ontology, regalia2018gnis, machado2010ontological, ping2009building}. This process is time intensive, requires domain expertise and is not generalizable.  Inspired by \cite{van2004method}, Janowicz and Ke{\ss}ler propose a framework that can be used to align a thesaurus with an ontology \cite{janowicz2008role}. The application of this framework is demonstrated by creating a place ontology using the ADL FTT. \cite{laurini2015geographic} presents a re-usable framework for multilingual place ontology alignment based on homologous relationships between entity toponyms, footprints, and types. They take a bottom up approach where they match the instances first in order to align the underlying ontology.  

Entity matching techniques detailed in the Appendix Table 1 can be used for automatic gazetteer alignment in a bottom up fashion where matches between entities or instances are used to integrate the broader concepts or feature types they belong to. \cite{brauner2007towards} manually annotate a dataset of matching features that are then used to integrate two thesauri based on the statistical matching frequency between feature type pairs. Similarly using annotated data, methods described in \cite{acheson2020machine,balsebre2022geospatial, hastings2008automated, martins2012using, martins2011supervised} use machine learning algorithms to align gazetteer data structures.  OWL:sameAs\footnote{https://www.w3.org/2001/sw/wiki/SameAs} links have also been exploited to align gazetteers [95, 110]. \cite{sunna2007structure} is an example of a structural approach to aligning place ontologies where the authors use a combination of concept level similarities as well as inheritance relations and relations between siblings. These automatic alignment methods are more generalizable and also require less domain expertise. When published as Linked Open Data, the methods can also lead to significant reduction of data redundancy, though they are often hindered by the absence of relations such as OWL:sameAs  and SKOS:exactmatch\footnote{https://www.w3.org/2009/08/skos-reference/skos.html} in ontology implementations \cite{delinked}.

\section{Volunteered Geographic Information}

The phrase Volunteered Geographic Information (VGI) was adopted in \cite{goodchild2007citizens} to refer to shared geographic information that was user generated, or crowdsourced, though crowdsourced geographic information was by then quite well established, for example with Wikimapia. Introduction of the term VGI coincided with the idea of non-authoritative, geographic knowledge systems that were self-governed by the public \cite{bishr2007geospatial}. VGI has become a major topic in the building of all geographic knowledge systems including gazetteers, and its presence permeates almost all aspects of gazetteer building, content, and maintenance. While some gazetteers are purely VGI-based there has been interest in the enrichment of authoritative gazetteers with VGI. Such efforts are however inevitably limited by the fact that the resulting gazetteer might no longer be classed as authoritative, and brings with it challenges of trust and data quality. There have however been numerous attempts to create new gazetteers integrating the two sources and incorporating the merits of both, like OSM and GeoNames.

\cite{craglia2012digital} propose a typology of VGI based on the way the information was made available by the community and the nature of the geographic information. If the data was made publicly available for a specific purpose it is treated as “explicit” otherwise it is  “implicit”. Similarly, if the information shared was explicitly geographic in nature, it is “explicit” and if the information is not about place but geographic information can be inferred, it is “implicit”.  Table 5 summarises gazetteers that use VGI in Craglia’s VGI typology \cite{craglia2012digital}.

\begin{table}[]
\centering 
\caption{Gazetteers created using intrinsic and extrinsic gazetteers}
\begin{tabular}{l|ll}
\hline
\multirow{2}{*}{}                                & \multicolumn{2}{c}{\textbf{Geography}}                                                         \\ \cline{2-3} 
                                                 & \multicolumn{1}{c|}{\textbf{Explicit}}                 & \multicolumn{1}{c}{\textbf{Implicit}} \\ \cline{2-3} 
\multirow{3}{*}{\textbf{Explicitly Volunteered}} & \multicolumn{1}{l|}{\textbf{OSM} \cite{beard2012semantic, gelernter2013automatic, lamprianidis2014extraction}}                      & \textbf{Wikipedia } \cite{rattenbury2007towards, popescu2009mining, gao2017constructing}                    \\
                                                 & \multicolumn{1}{l|}{\textbf{Panoramio} \cite{popescu2009mining}}                & \textbf{Dbpedia} \cite{lamprianidis2014extraction, moura2014integration}                      \\
                                                 & \multicolumn{1}{l|}{\textbf{Wikimpaia} \cite{gelernter2013automatic, lamprianidis2014extraction}}                & \textbf{Flickr} \cite{rattenbury2007towards, ahern2007world, popescu2009mining, gao2017constructing}                       \\ \hline
\multirow{4}{*}{\textbf{Implicitly Volunteered}} & \multicolumn{1}{l|}{\multirow{4}{*}{\textbf{Twitter} \cite{de2015leveraging}}} & \textbf{Gumtree} \cite{twaroch2008acquisition, twaroch2008mining}                      \\
                                                 & \multicolumn{1}{l|}{}                                  & \textbf{Phonebooks} \cite{goldberg2009extracting}                   \\
                                                 & \multicolumn{1}{l|}{}                                  & \textbf{Craigslist} \cite{hu2019natural}                  \\
                                                 & \multicolumn{1}{l|}{}                                  & \textbf{Search Engine results} \cite{jones2008modelling, popescu2009mining, pradeepa2016construction}        \\ \hline
\end{tabular}
\end{table}

Volunteered geographic information has great potential over authoritative sources in some contexts but is lacking in others. These advantages have been discussed in great detail in the literature \cite{liu2008search, mcdougall2009potential, rice2012supporting, fuchs2013tracing, el2019laying, daniel2023citizen, zahra2017natural} and can be summarised in the following three main points. 

\begin{itemize}
    \item VGI-based gazetteers are more accessible than authoritative gazetteers. VGI is almost always free and open source, not limited by restrictive licences of some authoritative gazetteers that require payments. This accessibility of VGI makes it attractive for students and researchers.
    \item VGI can be more up-to-date than authoritative sources. Updating a large authoritative knowledgebase is potentially a substantial task, often rolled out as a batch of small changes. In contrast, local changes in places may be noticed by individuals who can themselves update the VGI resource, where the changes might be regulated by other participants in the same area. The ability to make timely gazetteer (and associated map) updates at short notice is extremely helpful during disasters, requiring quick responses and situation awareness. 
    \item VGI can reflect local knowledge of individuals familiar with the locale, as opposed to being governed by a central body of experts with limited knowledge of local places. This enables VGI to capture vernacular and vague place names and identify their footprints with contextual information. Local knowledge can also give a high level of detail to features absent in authoritative sources. 
\end{itemize}

\subsection{Quality and Trust}

Drawing from \cite{bishr2007geospatial}, the authors of \cite{kessler2009agenda} focus on the challenges of integrating VGI in gazetteers and present a framework for trust in VGI. They argue that trust can be used as a proxy for the quality of VGI data and present a set of basic requirements for a trust model:

\begin{enumerate}
    \item Minimal metadata requirements from the user’s end.
    \item A feedback system must be in place that enables users to rate or assert the trust in the data obtained from the gazetteer.
    \item Computational models must be built to capture and compute a user’s reputation – a metric the authors define as the collective opinion of other system users on a particular user. 
    \item The trust model must be transparent allowing users to learn how someone becomes trustworthy.
    \item Provenance must be captured and integrated into the trust model – while identifying the difficulty in capturing provenance in VGI, authors suggest recording alternative data such as user’s interaction history with the system, trust ratings, cumulative user reputations, time stamps of contributions or modifications and user profile information.
\end{enumerate}

Several indicators have been explored in the literature for measuring the quality of VGI:

\begin{itemize}
    \item Credibility: credibility can be determined from the source of the piece of information and, like provenance, can be hard to capture in VGI. Expertise of the user is another measure of credibility \cite{flanagin2008credibility}.
    \item Local familiarity: The contributor’s familiarity with the locale can be used as an indication of their expertise in local knowledge \cite{senaratne2017review}.
    \item Experience: The user’s experience of the platform can be integrated in models that calculate the user’s reputation as discussed above \cite{van2010impact}. It can be reflected for example in the amount of time spent using the platform,  the volume of content created and or used, and the number of logins. 
    \item Reputation: The recognition of the user among other users. Indirect methods of calculating reputation would be to use past interactions of the user with other users, and the use of datasets published by the user. A rating system can be used to explicitly capture a user’s standing among peers \cite{maue2007reputation}.
    
\end{itemize}

Apart from the challenge of quality and accuracy of the data provided, VGI can also suffer from other disadvantages like sparsity and uneven coverage. The volume of volunteered data in VGI gazetteers has been seen to vary from continent to continent and country to country \cite{acheson2017quantitative}. While population sizes of countries can be seen as an obvious reason for this disparity, \cite{graham2015mapping} argue that other factors like a country’s population’s access to the internet and the state’s policies on open data can also have a significant impact on the coverage. \cite{acheson2017quantitative} also point out correlations between various feature types with the volume of information available – populated places had a strong correlation with feature counts when compared to natural feature types like mountains, hills and streams. 

Information directly obtained from VGI platforms or scraped from implicit crowd platforms contain most of the same components (geometry, feature type, temporality etc, see Section \ref {sec:Components}), required to construct or contribute to a gazetteer,  and face the same challenges that features from authoritative gazetteers encounter. Presentation of specific techniques to improve these VGI components is out of the scope of this paper. Interested readers can refer to the review of these quality measures and improvements in \cite{senaratne2017review}.

\section{Gazetteer Technologies}
\label{sec:GazTechs}

Early digital gazetteers that were digitizations of printed gazetteers were easily accommodated in databases such as relational database management systems (RDBMS). Earlier versions of the GNIS (Geographic Names Information System), Ordnance Survey gazetteer of the United Kingdom and the Alexandria Digital Library gazetteer are examples of gazetteers that used RDBMS.  Subsequent gazetteer technologies particularly in the last couple of decades have been dominated by the use either of extended (or object) relational databases, with their rich support for spatial data management, or of linked open data (LOD). Interest in the use of LOD was spurred by the appeal of creating openly accessible shared content that could be managed and queried using relatively standardised internet-based technologies. LOD uses semantic web technologies \cite{lassila2001semantic} that are designed to provide machine-readable content on the internet \cite{bernerslee2006linkeddata, bizer2011linked}. Linked open data is implemented with RDF (Resource Description Framework) that links entities or things identified by URIs (Uniform Resource Identifiers). RDF represents information as a set of triples, consisting of a subject, predicate, and object. 

\cite{kessler2009agenda} provide a possible stack of technologies that can be used to implement a distributed gazetteer. Even though the focus of that paper is mostly on integrating and building trust around VGI, their suggestion of using RDF triple stores with the SPARQL Protocol and RDF Query language (SPARQL) was reflected in  later gazetteers that used these technologies \cite{hara2017digital, hara2018digital}. SPARQL is the standard query language for RDF data stores and is used to query and manipulate data. Analogously with the development of SQL, it was extended to support geo-spatial data access in GeoSPARQL \cite{car2022geosparql, battle2012enabling}.

FODGS \cite{peng2010folksonomy} uses RDF and SPARQL with JENA-TDB, a scalable open source triplestore database by Apache. Titan, another open source distributed graph database management system that is built on top of Apache Cassandra, can be used as an RDF datastore \cite{moura2014integration}. Unlike JENA-TDB, Titan does not natively support SPARQL, but uses Gremlin as its query language. 

There are several examples of gazetteers that have used  OWL (Web Ontology Language) to build an ontology for the features, stored as RDF triplets, and queried with SPARQL/GeoSPARQL, \cite{beard2012semantic, cardoso2014gazetteer, cardoso2016swi}. OWL is one of the most widely used knowledge representation languages for representing and sharing ontologies on the World Wide Web \cite{gao2010design, arenas2013semantic, dang2011construction} and has gained W3C recommendation as a part of the Semantic Web technologies \cite{mcguinness2004owl}. Building on top of RDF and XML, OWL adds more vocabulary for describing properties and classes. It also adds relations between classes and richer typing of properties and characteristics. It should be noted that most of these gazetteers are designed to be available on the semantic web as linked open data and thus have interoperability as a top priority.

\begin{figure}[H]
\centering 
\includegraphics[width=0.8\textwidth]{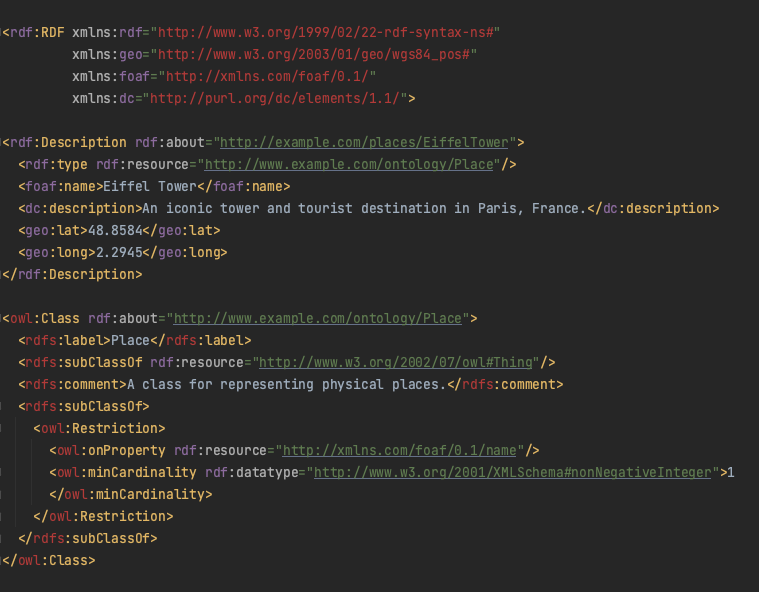}
    \caption[The system.]{Code snippet showing an example of a description of a place and its attributes and relations using RDF and OWL. }
\label{Fig5}
\end{figure}

The snippet from Fig. \ref{Fig5} represents a simple RDF graph that gives information about the Eiffel Tower. The namespaces are established in the first four lines. It then describes the RDF resource at http://example.com/places/EiffelTower. This is of type “Place”. It uses various tags from different namespaces to give this resource properties like name, description, and the location as a latitude and longitude pair. In the third section, the OWL class with the URI http://www.example.com/ontology/\\Place is described. Place is described as a subclass of a Thing and a restriction is imposed on the “name” property making the minimum cardinality for a place name to be 1.

Several advantages of storing gazetteer records as Linked Open Data can be identified:
\begin{enumerate}
    \item Interoperability: Linked open data can assist with integration and interoperability between different gazetteers and other datasets \cite{hara2017digital, cardoso2014gazetteer}.
    \item Data Quality:  By using open data standards and providing access to the data through a SPARQL endpoint, it is easier for other organisations and individuals to improve the data.
    \item Ease of Access: Linked open data makes it easier for end-users to access and query the data through tools such as semantic web browsers.
    \item Reusability: Linked open data allows for easy reuse of the data in different applications and domains, which can lead to new insights.
    \item Advanced Semantic searches: complex semantic searches can be performed when coupled with a powerful ontology (e.g. – OWL) \cite{janowicz2008role}.
    
\end{enumerate}

A recent effort aimed at building a common format for representing places is the Linked Places Format (LPF) developed jointly by the World Historical Gazetteer (WHG) project \cite{grossner2021linked} and the Pelagios project \cite{vitale2021pelagios}. The WHG uses the PostgreSQL/PostGIS relational database together with an Elastic search index for data storage. Elastic Search is an open-source search and analytics engine designed for real-time search functionality that enables full-text search. The LPF is built around JSON-LD (a format for structuring and representing linked data using JavaScript Object Notation) making it valid RDF and also valid GeoJSON (a JSON format for geographical features). They have enhanced the format enabling its recorded places to be scoped temporally with a “when” element through the use of GeoJSON-T \cite{ducatteeuw2021developing}. LPF does not intend to become the unified format or data model for records of places, but to be a format that facilitates linking between gazetteers. It helps record metadata that allows users to search across gazetteers, disambiguate and identify places and annotate data with stable URIs.

\begin{figure}[H]
\centering 
\includegraphics[width=0.9\textwidth]{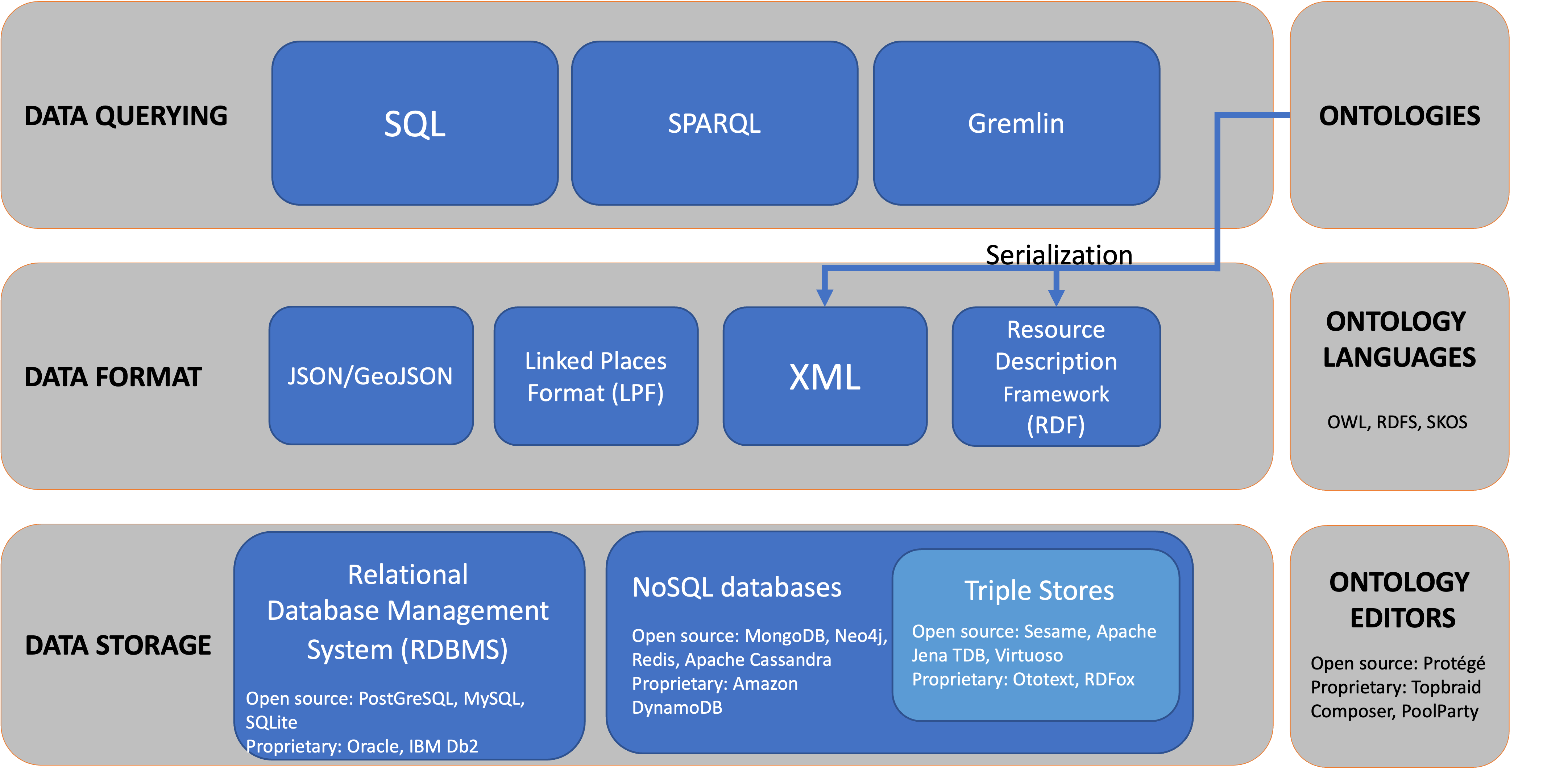}
    \caption[The system.]{Technology stack used to build modern digital gazetteers with some examples. }
\label{Fig6}
\end{figure}

PostGIS is an Object-Relational Database Management system that extends the open source PostgreSQL with support for geographic data storage and spatial query and is used in several gazetteers \cite{liu2009kidgs, rice2012supporting, adams2017wahi, ducatteeuw2021developing}. Its adoption may be attributed to the support provided for many spatial data formats such as Shapefile, KML, GeoJSON, WKT, as well as a wide range of spatial reference (coordinate) systems and transformations between them. It is also a natural development of earlier very widely used and familiar relational database technology. Fig. \ref{Fig6} shows some of these discussed technologies in a gazetteer building application. 

An issue that has been associated with the use of semantic web technologies for gazetteers is that many of the  published projects are associated with academic research projects and have not been complemented with support for longer term maintenance \cite{regalia2018gnis}.

As evident in the literature, the technologies used depend on the format of the published gazetteer. While most established gazetteers like GeoNames and OpenStreetMap have an LOD endpoint (both of them provide SPARQL endpoints), the question of what format to take depends on the use case of the gazetteer. If the gazetteer is expected to be used by expert users and is to be integrated, or interact, with other sources, or semantic searching is important, a linked data format is desirable. The downside of a LOD approach is the complexity of the task \cite{vitale2021pelagios} and the need to learn the technologies, which might be justified if the developers of the gazetteer are planning to maintain the gazetteer and the gazetteer is not a “means to an end but an end in itself” \cite{polczynski2022lessons}.

\section{Future Directions in Gazetteer Research}

In conducting this review several themes have emerged that might be the subject of future gazetteer research and development. These relate particularly to their level of information richness; multi temporality and the evolution of the recorded places; integration of the multiple representations of place in different gazetteers; trust and data quality; technologies and interoperability.

We have referred to the fact that the earliest, non-digital, gazetteers consisted of descriptions of places. While there can be many perspectives on the nature of a place, digital gazetteers have the potential to play a more effective role in providing more information rich records of place names that could be used to improve support for applications in information retrieval, particularly search for places with particular characteristics. A major limitation of most digital gazetteers is their failure to record much information about the nature of the place that is named, both its physical and social characteristics and, explicitly or implicitly, what are its affordances or uses. Often this does not extend beyond a single feature type. As was highlighted in the introduction there are many types of information that could be recorded about a named place. Analogously to, but going beyond, what is done in DBpedia and Wikipedia records of named places, it would be possible to introduce a set of properties relating for example to historical origin, political governance, commercial services (including for example eating and drinking places, banks, cinemas), industries, religious institutions, cultural institutions (museums, art galleries, theatres etc), natural features, parks, biodiversity, sports facilities, architectural styles, ethnic groups, notable inhabitants (and their dates), noise levels and crime statistics. It would also be possible to record personal experiences and perceptions based on volunteered information, as well as links to records in textual, audio and video documents relating to the place, including local and indigenous peoples' stories. The types of these data properties could be standardised to provide uniformity across gazetteers relating to different regions. 

Future gazetteers could be much richer with regard to the various formal and vernacular names in different languages, in association with their temporal and spatial extents. Given that spatial extent and location can change continuously for settlements and some natural features such as rivers and lakes, there is a role for functionality to retrieve these data for a point in time or to support spatio-temporal visualisation of the change over time of named places. Such functionality might require interacting with map databases and related historical data sources. It brings with it fundamental challenges of tracking the identity of a place which could change in association with factors such as the administrative regime, population level, ethnicity, industry, commerce, architecture, cultural activity and the natural environment. With the widespread availability of satellite imagery, latest developments in computer vision can also be incorporated more robustly, especially in the change of boundaries of geographic footprints. 

The current lack of consistency in the type and form of recorded information, in particular the use of feature type descriptors, or feature codes, hinders integration of multiple gazetteer sources that might refer at least partially to the same places. While it is reasonable to envisage encouraging the use of well-documented feature coding schemes, that cannot alleviate the need for methods to resolve multiple representations of the same place. The development of more accurate matching procedures is likely to be helped by the application of deep learning methods that encode words with embeddings that reflect their semantics and support determination of similarity between descriptors that could be lexically very different. In those respects, so-called blocking methods for filtering record matches have sometimes previously been quite inappropriate in using string similarity measures that could, for example, filter out equivalent names in different languages. Effective matching procedures would support the development of gazetteer portals - analogous to the meta-gazetteer concept - in which gazetteer records for a selected location could be retrieved and merged, or conflated, from multiple data sources, where those sources might differ in the types and richness of data recorded for a particular place. Instance level entity resolution methods, such as ones discussed in Section \ref{sec:integration} under de-duplication, are potent tools in matching multiple representations of the same place. Standardisation of these problems with standard datasets and metrics (as is the case with generic entity resolution)  would assist greatly in  enabling them to be compared with each other. Another requirement for solving this issue is the need for openly accessible datasets, as most of the latest methods rely on machine learning methods that are often supervised techniques that require training data. Though some of the methods in Appendix Table 1 were papers specifically publishing methods for entity resolution, only two of them made their datasets publicly available for replication or comparison. The challenge of representing or embedding heterogeneous geometry types (e.g. a point and a polygon or a line and a multi-polygon) in a deep-learning (neural) framework also remains unaddressed in methods published to date, as most deep-learning methods rely solely on point-point distance measures to compare footprints.

The common use of volunteered data as a gazetteer source is accompanied by lack of consistency and monitoring of data quality and hence trust. While the topic of VGI data quality has received considerable attention there remains scope for development of better methods to monitor and maintain it in gazetteers, for which it is possible to envisage training machine learning methods to infer measures of quality based on the associated metadata. 

There is a notable diversity in gazetteer data storage and management systems, exemplified by the contrast between relational databases and linked open data stores. Arguably attempts to try to dictate one type of data management technology are futile, but it could be beneficial to provide more widespread adoption of standard access interfaces that could be independent of the underlying storage technology. Progress has been made in this respect in the provision of APIs but there could still be benefit in providing standard forms of query as promoted by the Open Geospatial Consortium.

\section{Conclusions}

In reviewing the role of gazetteers as repositories of place name knowledge, we addressed the six research questions listed in the Introduction. (1) We described the evolution of gazetteers from historically recording the nature of individual places to storing a minimum of a few key components of name, geometric representation and feature type. (2) The source data of earlier digital gazetteers were often closely linked to the named content of digital maps, whether current or historical, while the content of some gazetteers has become dominated by volunteered content. Other gazetteers integrate names from authoritative map series with volunteered data while others are the result of merging some existing gazetteers. (3) When integrating gazetteer sources, one of the main challenges is that of determining whether two gazetteer records refer to the same real-world place. Automated methods for doing that have employed items of evidence based largely on the similarity of respectively, the place names, the feature types and the geometric footprints. Earlier automated methods were rule-based but current approaches usually employ machine learning including deep learning that transform the data items to be matched into word embeddings, or sometimes also geometric, embeddings. (4) We have described the increasingly important role of volunteered geographic information which in some cases dominates the content of a gazetteer. Its use is however accompanied by challenges of measuring and maintaining data quality. (5) When studying the implementation of gazetteers there is a clear distinction between those that are managed in spatially-enabled relational databases and those that are based on linked data technologies. At the time of writing it appears that relational databases are the dominant technology, as in widely used systems such as GeoNames and OpenStreetMap (with its gazetteer Nominatim) and for many national mapping agencies. Linked data technology is also well established to manage some gazetteer data both natively and to publish gazetteers implemented with other technologies. (6) We have identified significant limitations in digital gazetteers with regard to the lack of consensus on how data are represented, posing challenges in matching records between disparate sources. This relates particularly to feature types of which there are multiple classification and coding schemes. Word and geometry embeddings have potential to assist in harmonising representation. Development of improved data integration methods would benefit from much wider availability of data collections for training and testing.  A further prominent shortcoming is the sparsity of information about the nature of the named place. Much richer representation of place would assist in place-based information retrieval and cultural studies.  We also highlighted that knowledge-based gazetteers could record the spatio-temporal evolution of named places assisting in their role in historical information retrieval. 

In summary, gazetteers continue to play an essential role as place name knowledge resources that are of value in their own right but very importantly can be used to support information retrieval tasks that require the user to specify one or more place names. There is great potential for future research and development in the field, particularly with regard to providing much richer repositories of knowledge of places and of the ways in which named places have evolved over time and space and have changed in their physical, socio-economic and cultural nature.

\begin{acks}
The research was funded by the New Zealand Ministry of Business, Innovation and Employment (MBIE), grant number MAUX2104.
\end{acks}

\bibliographystyle{ACM-Reference-Format}
\bibliography{sample-base}


\appendix

\section{Appendix}

\begin{figure}[H]
\centering 
\includegraphics[width=0.5\textwidth]{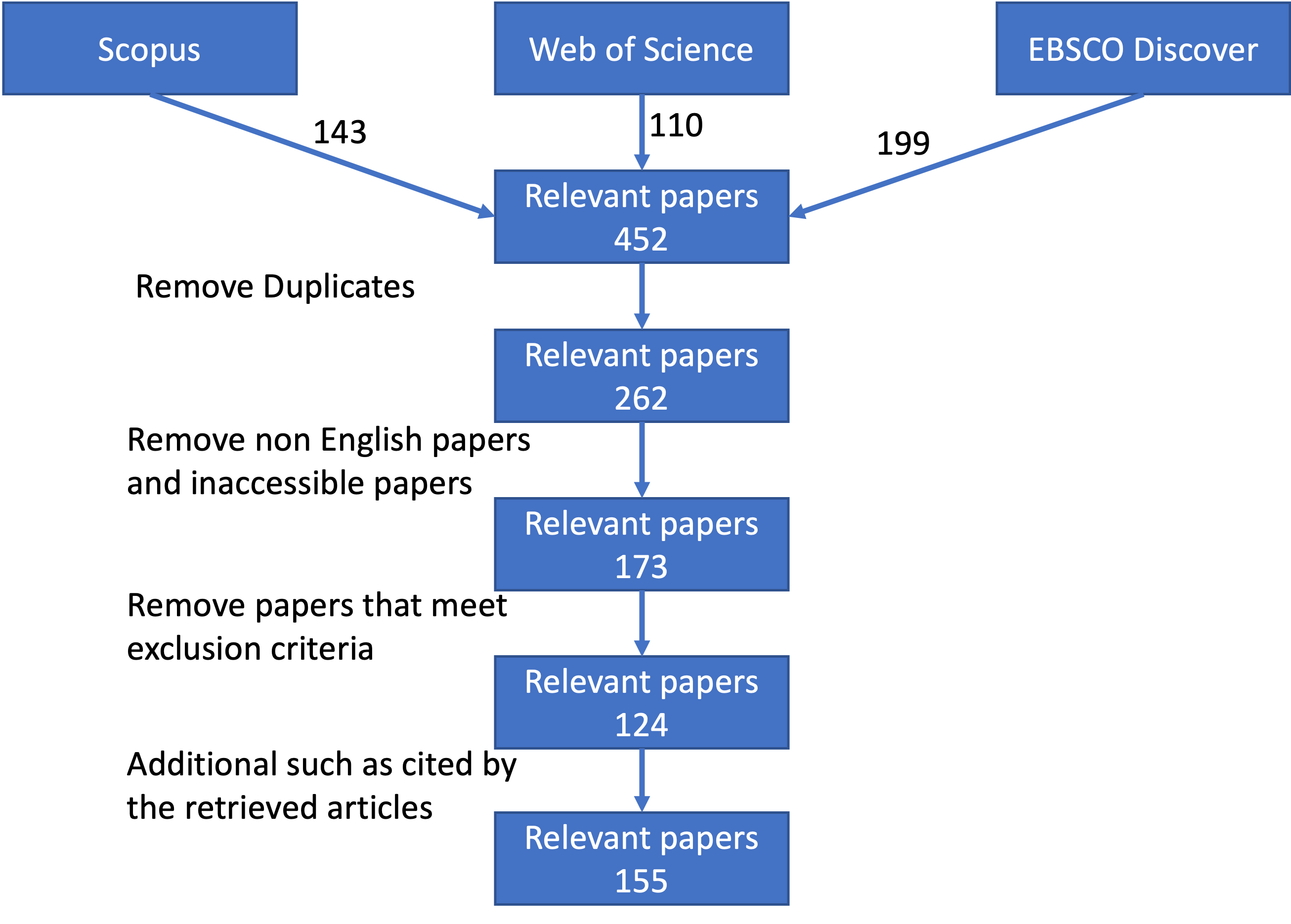}
    \caption[The system.]{Breakdown of the papers retrieved from the two search queries.}
\label{Fig1}
\end{figure} 

\begin{landscape}
\scriptsize

\begin{longtable}{|c|c|c|ccc|c|}
    \caption{Summary and comparison of papers that discuss duplicate identification and the methods used. Each number in the first column refers to a citation.  }
    \\\hline
\multirow{2}{*}{Paper} & \multirow{2}{*}{\begin{tabular}[c]{@{}c@{}}Duplicate\\ Identification\end{tabular}} & \multirow{2}{*}{\begin{tabular}[c]{@{}c@{}}Source/\\ Sources\end{tabular}}                                                                                                                                      & \multicolumn{3}{c|}{Components used for matching}                                                                                                                                                                                                                                                                                                                                                                                                                                                                                                                                                                                                                                        & \multirow{2}{*}{Notes}                                                                                                                                                                                                                                                                                     \\ \cline{4-6}
                       &                                                                                     &                                                                                                                                                                                                                 & \multicolumn{1}{c|}{Feature Name}                                                                                                                                                                                                    & \multicolumn{1}{c|}{\begin{tabular}[c]{@{}c@{}}Geographic \\ Footprint\end{tabular}}                                                                                                                                                     & Feature Type                                                                                                                                                                                           &                                                                                                                                                                                                                                                                                                            \\ \hline
\cite{hill1999indirect}                      & \begin{tabular}[c]{@{}c@{}}Inter-\\ gazetteer\end{tabular}                          & \begin{tabular}[c]{@{}c@{}}USGS Geographic\\ Name Service and\\ NIMA GeoNames\\ server\end{tabular}                                                                                                             & \multicolumn{1}{c|}{-}                                                                                                                                                                                                               & \multicolumn{1}{c|}{-}                                                                                                                                                                                                                   & -                                                                                                                                                                                                      & \begin{tabular}[c]{@{}c@{}}Uses GeoNames server \\ for places outside the \\ US and USGS \\ Geographic Names \\ Service for places \\ within the US therefore \\ avoiding duplicates\end{tabular}                                                                                                          \\ \hline
\cite{lucas2006information}                      & \begin{tabular}[c]{@{}c@{}}Inter-\\ gazetteer\end{tabular}                          & \begin{tabular}[c]{@{}c@{}}GeoNames server\\ and \\ GNIS\end{tabular}                                                                                                                                           & \multicolumn{1}{c|}{-}                                                                                                                                                                                                               & \multicolumn{1}{c|}{-}                                                                                                                                                                                                                   & -                                                                                                                                                                                                      & \begin{tabular}[c]{@{}c@{}}Uses GNS for places \\ outside of the US and \\ GNIS for places within \\ the US therefore \\ avoiding duplicates\end{tabular}                                                                                                                                                  \\ \hline
\cite{sehgal2006entity}                      & \begin{tabular}[c]{@{}c@{}}Inter and \\ Intra \\ gazetteer\end{tabular}             & \begin{tabular}[c]{@{}c@{}}United\\ Kingdom Permanent\\ Committee on\\ Geographic Names \\ and Geographic\\  Names Database of\\  the United States \\ Board on\\ Geographic Names\end{tabular}                 & \multicolumn{1}{c|}{\begin{tabular}[c]{@{}c@{}}Edit distance, \\ Jaccard and \\ Jaro-Winkler \\ similarity \\ measures\end{tabular}}                                                                                                 & \multicolumn{1}{c|}{\begin{tabular}[c]{@{}c@{}}Inverse geographic \\ distance between \\ coordinate pairs\end{tabular}}                                                                                                                  & \begin{tabular}[c]{@{}c@{}}Type similarity \\ based on co-\\ occurrence in the \\ manually annotated \\ dataset they have \\ prepared\end{tabular}                                                     & \begin{tabular}[c]{@{}c@{}}Features are fed into \\ logistic regression, \\ voted perceptron \\ (Neural Network) and \\ support vector\\  machine models. \\ Logistic regression \\ achieves best results\end{tabular}                                                                                     \\ \hline
\cite{vestavik2007merging}                      & \begin{tabular}[c]{@{}c@{}}Inter-\\ gazetteer\end{tabular}                          & \begin{tabular}[c]{@{}c@{}}ADL and Sentralt \\ Stedsnavnregister \\ (SSR) (Official \\ Norwegian place \\ registry)\end{tabular}                                                                                & \multicolumn{1}{c|}{\begin{tabular}[c]{@{}c@{}}Number of \\ shared place \\ names\end{tabular}}                                                                                                                                      & \multicolumn{1}{c|}{\begin{tabular}[c]{@{}c@{}}Proximity and \\ whether contained \\ by same larger \\ feature\end{tabular}}                                                                                                             & \begin{tabular}[c]{@{}c@{}}Hand crafted \\ mapping between \\ two thesauri. \\ Similarity measured \\ as a graph distance \\ over two \\ vocabularies.\end{tabular}                                    &                                                                                                                                                                                                                                                                                                            \\ \hline
\cite{hastings2008automated}                      & \begin{tabular}[c]{@{}c@{}}Inter and \\ Intra\\  gazetteer\end{tabular}             & \begin{tabular}[c]{@{}c@{}}Geographic Data \\ Technology (GDT, \\ now TeleAtlas), \\ GNIS, ESRI ‘Data and \\ Maps CD’, \\ ‘Transportation 2.0’ \\ data suite, Lake \\ Tahoe Data Clearing \\ house\end{tabular} & \multicolumn{1}{c|}{\begin{tabular}[c]{@{}c@{}}A stemming \\ string similarity \\ algorithm\end{tabular}}                                                                                                                            & \multicolumn{1}{c|}{\begin{tabular}[c]{@{}c@{}}Collocation used as a \\ threshold for candidate \\ selection. Area of \\ overlap of footprints \\ used for matching\end{tabular}}                                                        & \begin{tabular}[c]{@{}c@{}}The steps to the \\ common ancestor \\ in the feature type \\ hierarchy\end{tabular}                                                                                        & \begin{tabular}[c]{@{}c@{}}Proposes methods for \\ duplicate identification. \\ Demonstrates with the \\ mentioned sources. \\ Matching of different \\ thesauri is out of scope\end{tabular}                                                                                                              \\ \hline
\cite{manguinhas2008geo}                      & \begin{tabular}[c]{@{}c@{}}Inter-\\ gazetteer\end{tabular}                          & \begin{tabular}[c]{@{}c@{}}Geonames, \\ GeoNetPT, \\ Google Earth \\ Community, Estonian \\ National Library,\\  ECAI time period \\ directory, Wikipedia\end{tabular}                             & \multicolumn{1}{c|}{\begin{tabular}[c]{@{}c@{}}Manually tuned \\ threshold for \\ Jaro-Winkler \\ token similarity\end{tabular}}                                                                                                     & \multicolumn{1}{c|}{\begin{tabular}[c]{@{}c@{}}Distance below \\ a manually tuned \\ threshold\end{tabular}}                                                                                                                             & \multicolumn{1}{c|}{\begin{tabular}[c]{@{}c@{}}Introduce a novel \\ ontology which \\ features from \\ all sources \\ are manually \\ mapped to\end{tabular}}                                                                                                                                                                                                              & \begin{tabular}[c]{@{}c@{}}-\end{tabular}                                                                                                                                                                                                  \\ \hline
\cite{cheng2010data}                      & \begin{tabular}[c]{@{}c@{}}Inter and \\ intra \\ gazetteer\end{tabular}             & \begin{tabular}[c]{@{}c@{}}National Geographical \\Names Database and \\geographical names \\information system \\issued by Chinese \\Ministry of Civil \\affairs\end{tabular}                                                                                                                                                                                                         & \multicolumn{1}{c|}{\begin{tabular}[c]{@{}c@{}}Threshold for \\ token similarity\end{tabular}}                                                                                                                                       & \multicolumn{1}{c|}{\begin{tabular}[c]{@{}c@{}}Distance below \\ a threshold \\ containment\end{tabular}}                                                                                                                                & \begin{tabular}[c]{@{}c@{}}Manually maps \\ sources to a \\ reference ontology\\ and obtains \\ a semantic \\ similarity\end{tabular}                                                                                                      & \begin{tabular}[c]{@{}c@{}}Uses a weighted sum \\ of the three types of \\ similarities similar \\ to Hastings, 2008.\end{tabular}                                                                                                                                                                         \\ \hline
\cite{zheng2010detecting}                      & \begin{tabular}[c]{@{}c@{}}Intra-\\ gazetteer\end{tabular}                          & Bing Maps                                                                                                                                                                                                       & \multicolumn{1}{c|}{\begin{tabular}[c]{@{}c@{}}Edit distances \\ and inverse \\ document \\ frequency \\ of similar and \\ dissimilar strings \\ between names\end{tabular}}                                                         & \multicolumn{1}{c|}{\begin{tabular}[c]{@{}c@{}}Uses address instead \\ of a footprint. \\ Builds an address \\ hierarchy and \\ calculates distance \\ to coparent.\end{tabular}}                                                        & \begin{tabular}[c]{@{}c@{}}The levels to reach \\ the coparent for \\ the two features \\ in the category \\ hierarchy\end{tabular}                                                                    & \begin{tabular}[c]{@{}c@{}}Uses a decision tree\\ classifier for matching\end{tabular}                                                                                                                                                                                                                     \\ \hline
\cite{smart2010multi}                      & \begin{tabular}[c]{@{}c@{}}Inter-\\ gazetteer\end{tabular}                          & \begin{tabular}[c]{@{}c@{}}GeoNames, OSM, \\ Yahoo WOE, \\ Wikipedia, \\ OS Point X, \\ OS 50k, \\ OS MasterMap\end{tabular}                                                                                    & \multicolumn{1}{c|}{\begin{tabular}[c]{@{}c@{}}Levenshtein \\ edit distance, \\ Soundex, \\ Text \\ normalisation\end{tabular}}                                                                                                      & \multicolumn{1}{c|}{\begin{tabular}[c]{@{}c@{}}Distance threshold \\ applied to \\ distances between \\ geometry objects of \\ points and polygons\end{tabular}}                                                                         &                                                                                                                                                                                                        & \begin{tabular}[c]{@{}c@{}}A threshold is applied to \\ a weighted combination \\ of Levenshtein and \\ Soundex distances \\ before applying \\ a spatial distance \\ threshold.\end{tabular}                                                                                                              \\ \hline
\cite{martins2011supervised}                     & \begin{tabular}[c]{@{}c@{}}Intra-\\ gazetteer\end{tabular}                          & \begin{tabular}[c]{@{}c@{}}ADL and other \\unspecified sources\end{tabular}                                                                                                                                                                                                                 & \multicolumn{1}{c|}{\begin{tabular}[c]{@{}c@{}}Variety of string \\ similarity features \\ including \\ Levenshtein,\\  Jaro-Winkler \\ and Monge-Elkan \\ distances\end{tabular}}                                                   & \multicolumn{1}{c|}{\begin{tabular}[c]{@{}c@{}}Variety of distance \\ (minimum distance \\ between footprints, \\ distance between \\ centroids, etc), \\ overlap of features,\\  overlap relative to\\  feature size, etc\end{tabular}} & \begin{tabular}[c]{@{}c@{}}Variety of features \\ like Jaccard \\ coefficient, Dice \\ coefficient, etc \\ of place types \\ associated with \\ the place, up-steps \\ to common ancestor\end{tabular} & \begin{tabular}[c]{@{}c@{}}Features fed into an SVM \\ and alternating decision \\ tree classifier. \\ Author has also \\ considered semantic \\ relations and \\ temporal features.\\ Method proposed \\ however is\\ only applicable to \\ features from the same \\ feature type thesauri\end{tabular} \\ \hline
\cite{gelernter2013automatic}                     & \begin{tabular}[c]{@{}c@{}}Inter-\\ gazetteer\end{tabular}                          & \begin{tabular}[c]{@{}c@{}}GeoNames, \\ Wikimapia, \\ OpenStreetMap\end{tabular}                                                                                                                                & \multicolumn{1}{c|}{\begin{tabular}[c]{@{}c@{}}Weighted N-gram \\ similarity, edit \\ distance, \\ soundex \\ matching scores\end{tabular}}                                                                                          & \multicolumn{1}{c|}{\begin{tabular}[c]{@{}c@{}}Normalized \\ geographic \\ distance\end{tabular}}                                                                                                                                        & -                                                                                                                                                                                                      & \begin{tabular}[c]{@{}c@{}}Uses two SVMs to \\ learn weights for the \\ n-gram queries for \\ place names and \\ to rank the candidates. \\ Gives a fuzzy result.\end{tabular}                                                                                                                             \\ \hline
\cite{moura2014integration}                     & \begin{tabular}[c]{@{}c@{}}Inter-\\ gazetteer\end{tabular}                          & \begin{tabular}[c]{@{}c@{}}GeoNames, \\ DBPedia\end{tabular}                                                                                                                                                    & \multicolumn{1}{c|}{\begin{tabular}[c]{@{}c@{}}Only exact name \\ matches are \\ considered\end{tabular}}                                                                                                                            & \multicolumn{1}{c|}{\begin{tabular}[c]{@{}c@{}}containedBy predicate\\  is used to check \\ common places that \\ contain the \\ two candidates\end{tabular}}                                                                            & -                                                                                                                                                                                                      & \begin{tabular}[c]{@{}c@{}}Exploit semantic tags \\ and Wikipedia links. \\ The sameAs predicate \\ or the sharing of a \\ common wikipedia link \\ is considered to be an \\ exact match between the \\ two linked data entries.\end{tabular}                                                             \\ \hline
\cite{isaj2019multi}                     & \begin{tabular}[c]{@{}c@{}}Inter and \\ Intra POI\end{tabular}                      & \begin{tabular}[c]{@{}c@{}}Google Places, \\ Foursquare, \\ Yelp, and Krak\end{tabular}                                                                                                                         & \multicolumn{1}{c|}{\begin{tabular}[c]{@{}c@{}}Levenshtein \\ string \\ similarity\end{tabular}}                                                                                                                                     & \multicolumn{1}{c|}{\begin{tabular}[c]{@{}c@{}}Filtering with \\ an adapted \\ quadtree\end{tabular}}                                                                                                                                    & \begin{tabular}[c]{@{}c@{}}Wu-Palmer \\ similarity applied \\ to pairs of \\ attributes from \\ different entities\end{tabular}                                                                        & \begin{tabular}[c]{@{}c@{}}Non-supervised pareto \\ optimisation applied to \\ text and semantic \\ similarity measures in \\ combination with \\ determining threshold \\ for skyline to distinguish \\ positive / negative matches.\end{tabular}                                                         \\ \hline
\cite{acheson2020machine}                     & \begin{tabular}[c]{@{}c@{}}Inter-\\ Gazetteer\end{tabular}                          & \begin{tabular}[c]{@{}c@{}}Geonames, \\ SwissNames3D\end{tabular}                                                                                                                                               & \multicolumn{1}{c|}{\begin{tabular}[c]{@{}c@{}}Levenshtein-\\ Damerau \\ distance, the \\ Jaro similarity, \\ and the \\ Jaro-Winkler \\ similarity.\\  Also uses \\ Levenshtein \\ distance on any \\ alternate names\end{tabular}} & \multicolumn{1}{c|}{\begin{tabular}[c]{@{}c@{}}point-to-point \\ distance between \\ gazetteer records\end{tabular}}                                                                                                                     & \begin{tabular}[c]{@{}c@{}}one-hot encoding \\ to encode \\ feature types\end{tabular}                                                                                                                 & \begin{tabular}[c]{@{}c@{}}Features fed into \\ a random forest \\ classifier to classify \\ pairs of places. \\ Also use other \\ attributes like elevation \\ and land cover. \\Also compares results\\with a rule based \\approach\end{tabular}                                                                                                                \\ \hline
\cite{balsebre2022geospatial}                     & \begin{tabular}[c]{@{}c@{}}Inter-\\ Gazetteer\end{tabular}                          & \begin{tabular}[c]{@{}c@{}}Yelp, \\ Foursquare \\ and OSM\end{tabular}                                                                                                                                          & \multicolumn{1}{c|}{\begin{tabular}[c]{@{}c@{}}BERT embedding \\ of text\end{tabular}}                                                                                                                                               & \multicolumn{1}{c|}{\begin{tabular}[c]{@{}c@{}}A distance \\ embedding that is \\ a function of\\ Haversine distance\end{tabular}}                                                                                                       & \begin{tabular}[c]{@{}c@{}} - \end{tabular}                                                                                                                                      & \begin{tabular}[c]{@{}c@{}}A blocking phase selects\\ pairs that meet a string\\ similarity and distance\\ threshold. Also uses a\\ contextual embedding\\ based on combining\\ BERT embeddings of\\ target entities with \\ nearby entities. All\\ embeddings combined to\\ make predictions\end{tabular} \\ \hline

\label{ERTable}
\end{longtable}
\end{landscape}

\end{document}